\documentclass[aps,onecolumn,superscriptaddress,showpacs]{revtex4}
\usepackage{epsfig,graphicx}
\usepackage{indentfirst}
\usepackage{amsfonts,amssymb,latexsym,amsmath,enumerate,amsthm}
\usepackage{bm}
\usepackage{color}
\def\rot{{\rm rot}}

\def\Im{{\rm Im}}

\def\sgn{{\rm sgn}}
\newcommand{\be}{\begin{equation}}
\newcommand{\ee}{\end{equation}}
\newcommand{\bea}{\begin{eqnarray}}
\newcommand{\eea}{\end{eqnarray}}
\newcommand{\dst}{\displaystyle}
\newcommand{\fr}[2]{\frac{{\dst #1}}{{\dst #2}}}

\begin{document}

\title{Polarization radiation of vortex electrons with large orbital angular momentum}

\author{Igor~P.~Ivanov}
    \email[E-mail: ]{igor.ivanov@ulg.ac.be}
\affiliation{IFPA, Universit\'{e} de Li\`{e}ge, All\'{e}e du
6 Ao\^{u}t 17, b\^{a}timent B5a, 4000 Li\`{e}ge, Belgium}
\affiliation{Sobolev Institute of Mathematics, Koptyug avenue 4, 630090, Novosibirsk, Russia}
\author{Dmitry~V.~Karlovets}
    \email[E-mail: ]{d.karlovets@gmail.com}
\affiliation{Tomsk Polytechnic University,
Lenina 30, 634050 Tomsk, Russia}

\date{\today}

\begin{abstract}
Vortex electrons, --- 
freely propagating electrons whose wavefunction has helical wavefronts, --- could become a novel tool in the physics  
of electromagnetic radiation. They carry a non-zero intrinsic orbital angular momentum (OAM) $\ell$ with respect to the propagation axis
and, for $\ell \gg 1$, a large OAM-induced magnetic moment, $\mu \approx \ell \mu_B$ ($\mu_B$ is the Bohr magneton),
which influences the radiation of electromagnetic waves. 
Here, we consider in detail the OAM-induced effects by such electrons in two forms of polarization radiation, namely in Cherenkov radiation and transition radiation. 
Thanks to the large $\ell$, 
we can neglect quantum or spin-induced effects, which are of the order of $\hbar \omega/E_e \ll 1$, but retain the magnetic moment contribution $\ell \hbar \omega/E_e \lesssim 1$, which makes the quasiclassical approach to polarization radiation applicable. 
We discuss the magnetic moment contribution to polarization radiation, which has never been experimentally observed, and study how its visibility depends on the kinematical parameters and the medium permittivity. In particular, it is shown that this contribution can, in principle, be detected in azimuthally non-symmetrical problems, for example when vortex electrons obliquely cross a metallic screen (transition radiation) or move nearby it (diffraction radiation). 
We predict a left-right angular asymmetry of the transition radiation (in the plane where the charge radiation distributions would stay symmetric), which appears due to an effective interference between the charge radiation field and the magnetic moment one. Numerical values of this asymmetry for vortex electrons with $E_e = 300$ keV and $\ell = {\cal O}(100-1000)$ are ${\cal O} (0.1-1\%)$, and we argue that this effect could be detected with existing technology. The finite conductivity of the target and frequency dispersion play the crucial roles in these predictions.

\end{abstract}


\maketitle

\section{Introduction}

Radiation of electromagnetic (EM) waves is an inherent property of charges.
In electrodynamics, there exist two general classes of radiation: bremsstrahlung and polarization radiation (PR).
Bremsstrahlung is produced by a charge accelerated in some external EM field, and it comprises such processes as synchrotron radiation, undulator radiation, bremsstrahlung in a Coulomb field, etc. In contrast, there are various forms of PR, such as Cherenkov radiation, transition radiation,
diffraction radiation, Smith-Purcell radiation, parametric X-ray radiation, etc., which can be emitted by a uniformly moving charge
but only in the presence of a medium. In this case, at each point of the medium the time-varying EM field of the moving particle
induces time-varying currents, which are sometimes called the polarization currents and may be considered as a radiation source (see e.g., \cite{Ryaz, DR, Amusia, JETP}).
In a microscopic treatment, PR arises as a result of the so-called distant collisions of a particle with an atom or molecule. In this case, the effective (mainly) dipole moments induced by the projectile's field inside the target emit only soft photons, and the particle trajectory stays undisturbed, see e.g., \cite{Amusia, Ryaz, DR}. 

It is clear that EM radiation can be produced not only by charges but also by neutral particles
carrying higher multipoles: electric or magnetic dipoles, quadrupoles, etc.
For example, there is a vast literature on the problem of the spin magnetic moment radiation in external fields 
and in matter (the so-called ``spin light'') for electrons, neutrinos, etc. (see e.g., \cite{Light, Neutrino, Sakuda, Gover} and the references therein). 
Then, Cherenkov radiation by a neutron treated as a pointlike particle with a zero charge but with a magnetic dipole moment 
is a well-known problem (see e.g., \cite{Frank, GTs}).
Transition radiation by the magnetic moments as well as the electric dipoles and quadrupoles also have been analyzed in detail in \cite{GTs}.

It is therefore remarkable that despite a big theoretical interest, the experimental observations of 
the magnetic moment (or any higher multipole) influence on the EM radiation are very scarce.
Putting aside various spin-dependent radiative processes in high-energy particle collisions,
they are, in fact, limited to only very few cases of the bremsstrahlung of ultrarelativistic electrons.
For example, in \cite{NIMR1984}, the synchrotron radiation intensity in the $100-400$ keV range 
at the VEPP-$4$ storage ring for $5$ GeV electrons was found to depend on the electron spin orientation. 
The effect magnitude was small, of the order of $10^{-4}$--$10^{-3}$,
but due to the high photon counting statistics it was well measurable.
This effect was even proposed as a tool for measuring the beam transverse polarization at storage rings.
Further development of this idea led to a proposal of a ``spin-light polarimeter'' for the future $12$ GeV JLab storage ring \cite{Dutta}.
Spin effects in bremsstrahlung were also observed at CERN by detecting the GeV-range photons emitted by the $35-243$ GeV electrons passing through a tungsten single crystal 
(out of the channeling regime) \cite{Br}. The effect was detectable due to 
the super-strong EM field in the crystal comparable to the Sauter-Schwinger limit in the electron rest frame.

In contrast to these results for bremsstrahlung, the magnetic moment (or any higher multipole) contribution to any form of {\em polarization radiation} 
has never been detected. There are several obstacles to this measurement. On the purely experimental side, a ``no-win'' situation: 
the PR intensity is, roughly speaking,
larger for soft photons, especially in the coherent regime of emission (see e.g., \cite{Gover}), but the relative contribution of the spin-induced magnetic moment is attenuated by $\hbar \omega/E_e$, where $\hbar \omega$ and $E_e$ are the photon energy and the electron energy, respectively. 
However even putting aside this experimental difficulty, there is a deeper problem of separating the spin-induced magnetic moment contribution to PR
from the quantum recoil effects, which are of the same order (this fact was ignored in the analysis of Ref.~\cite{Gover}).
Indeed, in the macroscopic quasiclassical treatment of PR, one assumes that the particle trajectory stays unperturbed by the radiation.
In other words, one neglects the effects of the order of $\hbar \omega/E_e$ from the very beginning,
and the spin-induced magnetic moment contribution to PR lies beyond the standard calculation scheme. 
As for the quantum theory of PR, which is far from being completed as yet (see e.g., \cite{TM, G}), the spin magnetic moment contribution, again, 
has the quantum recoil effects as a natural competitor, which makes an experimental separation of both contributions a rather delicate task.

The theoretical prediction \cite{Bliokh-07} and the recent experimental demonstration of the vortex electron beams \cite{vortex-experimental-tonomura,vortex-experimental-verbeeck,vortex-experimental-mcmorran,vortex-experimental-uchida} put a dramatic twist on this problem.
Vortex electrons carry an intrinsic orbital angular momentum (OAM) $L = \hbar \ell$ with respect to their average propagation direction, 
and the values of $\ell$ can be rather large (up to $100$ in \cite{vortex-experimental-mcmorran} and up to $90$ in \cite{vortex-experimental-uchida}). The magnetic moment associated with the OAM is correspondingly large \cite{Bliokh-11}, $\mu \approx \ell \mu_B$, where $\mu_B = e\hbar/2 m_e c$ is the Bohr magneton. 
It strongly enhances all the magnetic moment effects compared to the usual spin contribution, $2\mu_B$.
Using vortex electrons with $\ell \gg 1$, one can enter the regime in which the magnetic-moment contribution is only moderately suppressed, $\propto\ell \hbar \omega/E_e \lesssim 1$, and it remains much larger than the quantum effects. 
This improves the visibility of the magnetic moment contribution 
to PR and, at the same time, makes its quasiclassical calculation a selfconsistent problem.
An observation of this contribution would be the first clear evidence of the PR by a multipole. 

As a particular example, we considered in \cite{IK-short} transition radiation of vortex electrons with $\ell \gg 1$ obliquely incident on a metallic foil
and predicted that the OAM-induced magnetic moment contribution could manifest itself via a left-right asymmetry of the radiation.
For electrons with $E_e = 300$ keV, which is a typical energy of the vortex electrons in electron microscopes,
and $\ell \sim{\cal O}(1000)$, the asymmetry magnitude can be of the order of ${\cal O}(1\%)$, 
which must be well detectable. In this paper, we present a fuller discussion of this process, including its dependence on the kinematical parameters 
and on the medium permittivity, $\varepsilon(\omega)$,
as well as a comparison with Cherenkov radiation by vortex electrons.

The structure of the paper is as follows. In Sec.~\ref{section-qualitative} we remind the reader of the qualitative features of transition radiation 
from a charge and a magnetic moment. We then pass to an accurate description of the transition radiation from a system ``charge + magnetic moment'' and present in Sec.~\ref{section-formulas} the formulas
for two quasiclassical ways of modelling the magnetic moment. The numerical results are given in Sec.~\ref{section-numerical}. In Sec.~\ref{section-discussion} we discuss the results and outline the requirements and a strategy to detect the proposed effect in an experiment.

\section{Transition radiation from ``charge + magnetic dipole'': Qualitative features} 
\label{section-qualitative}

\subsection{General properties of PR }

Polarization radiation (PR) occurs when a particle moves uniformly near or inside a medium with the complex permittivity $\varepsilon (\omega) = \varepsilon^{\prime} + i\varepsilon^{\prime \prime}$ \footnote{As we shall be interested mostly in the optical and UV spectral regions, one can set the magnetic permeability to unity; this condition is justified for the majority of real substances (see e.g., \cite{L8}).}. 
Depending on the medium or target shape, one usually distinguishes different particular types of PR:
Cherenkov radiation (ChR), transition radiation (TR), diffraction radiation (DR), Smith-Purcell radiation (SPR), parametric X-ray radiation, etc. (see e.g., \cite{DR, Amusia, JETP}).
Along with the energy losses to excitation and ionization of the atomic shells, which result in a discrete spectrum radiation of the relatively \textit{hard} photons (bremsstrahlung), there are also the so-called polarization losses related to the dipole moments induced inside the medium and leading to a continuous spectrum radiation of the relatively \textit{soft} photons (see e.g., \cite{Amusia} and the references therein). Although many \textit{macroscopic} manifestations of PR were known
since $1930$-$50$'s (ChR, TR, DR, SPR), the \textit{microscopic}  quantum theory of PR explicitly demonstrating their common physical origin was developed only in the $1970$-$80$'s by Amusia with co-workers (see e.g., \cite{Amusia, Amusia_76} and the references therein; qualitative explanations of the microscopic nature of, say, ChR were of course given before). The macroscopic approaches, in which such a unified nature of various radiation processes was explicitly demonstrated, have been developed only in recent years \cite{Sysh, Tish, DR, JETP}.

As a matter of fact, radiation of soft photons (the ones for which $\omega \ll E_e$) represents a somewhat complementary process to the usual bremsstrahlung of an accelerated charge, since, as we know, only the \textit{sum} of probabilities of these two processes is measured in experiment. 
One of the most remarkable differences between the ordinary bremsstrahlung and PR is that whereas intensity of the former is inversely proportional to the projectile (say, electron) mass squared, $dW\propto m_e^{-2}$, the intensity of the latter has no dependence on this mass at all. As a result, PR can even dominate over bremsstrahlung, especially in the ultrarelativistic case \cite{DR, Amusia}.

Due to the different kinematic conditions, various types of PR have different spectra, but the shape of the latter, nevertheless, is mostly defined by the permittivity dispersion. In particular, in the ultrarelativistic case the spectrum (say, of TR \cite{GTs, G}) can span up to the frequencies $\omega_c \sim \gamma \omega_p$ ($\gamma = E_e/m_ec^2 = 1/\sqrt{1-\beta^2}$ is the Lorentz factor), which 
 can lie in the X-ray region for very energetic electrons, since the plasma frequency $\omega_p$ is around $10-30$ eV for many materials.

\subsection{General properties of charge TR }

One of the simplest and widely known types of PR is transition radiation, which occurs when a uniformly moving charge crosses an interface separating two media with different permittivities.
Put simply, although the charge motion is uniform, the accompanying EM field reorganizes itself when crossing the interface, and it
is partly ``shaken off'' in the form of EM radiation. The simplest example of TR at the normal incidence was considered in the seminal paper 
by Ginzburg and Frank \cite{Ginzburg-Frank}.
In the following decades, a theory treating the physics of transition radiation in ever increasing details and in more general set-ups has gradually emerged (see e.g., Ref.~\cite{Paf} and also the monograph \cite{GTs}) and even has become a standard textbook material \cite{L8}.
There are several aspects which enrich the phenomenon of TR and complicate its theoretical investigation: normal vs. oblique incidence,
an ideal conductor vs. a medium with an arbitrary complex permittivity $\varepsilon$, one interface vs. multiple interfaces (see e.g., \cite{Paf}), etc. 

\begin{figure}[!htb]
   \centering
\includegraphics[width=6cm]{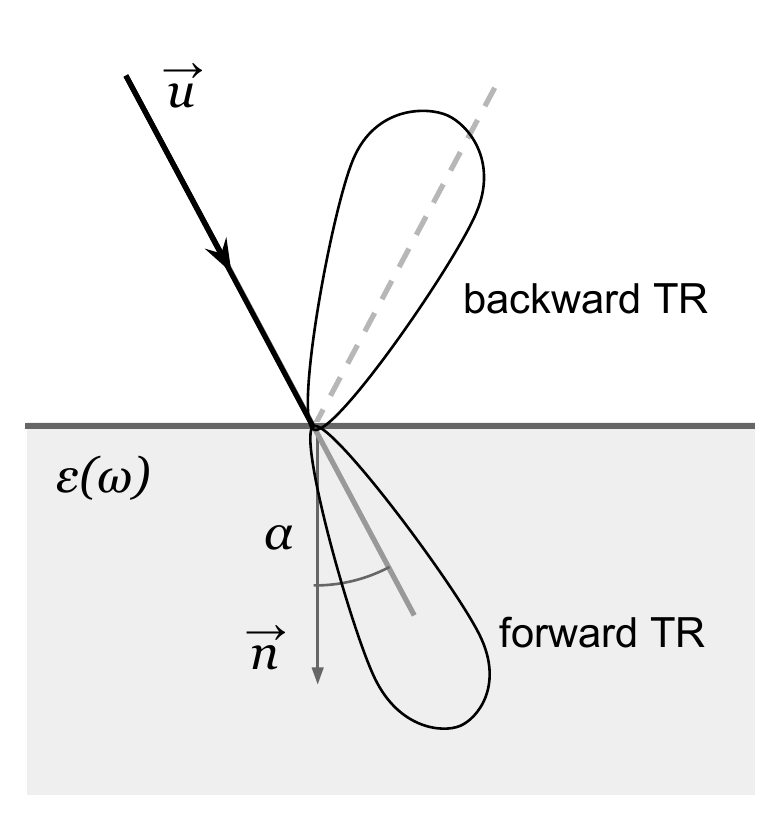}\hspace{1cm}
\includegraphics[width=7cm]{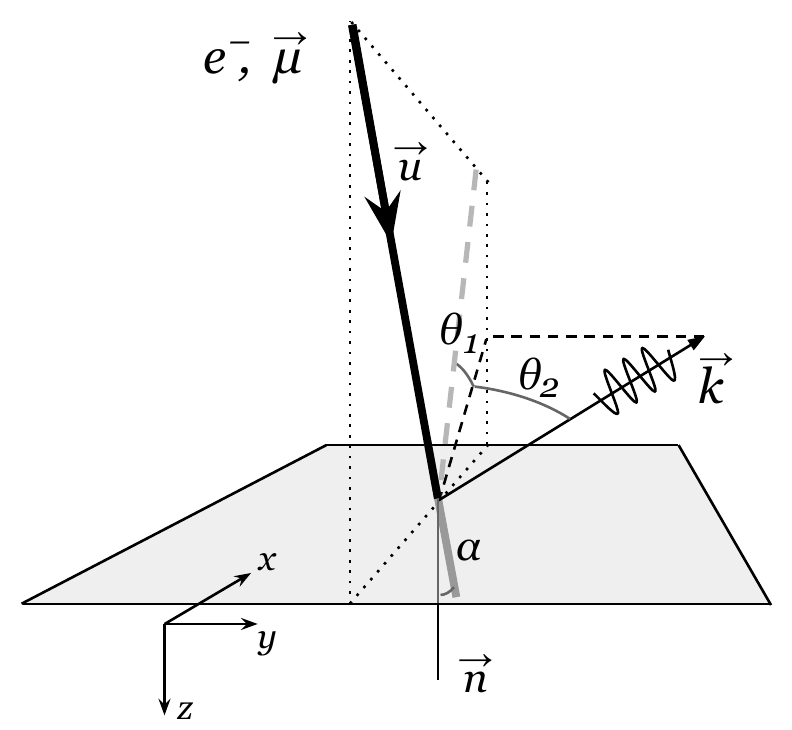}
\caption{(Left) A schematical view of the forward and backward TR lobes projected onto the incidence plane. (Right) Angle conventions for an oblique incidence with the example of the backward TR. The specular reflection direction is shown by the gray dashed line.}
   \label{fig-geometry}
\end{figure}

One of the specific features of TR is the radiation generation to both semispaces: the one the particle is coming from 
(backward TR) and the one the particle enters after crossing the interface (forward TR), see Fig.~\ref{fig-geometry}, left. 
For the case of a vacuum-ideal conductor interface, this can illustratively be explained as a radiation from two charges which annihilate at the interface 
(in order to comply with the boundary conditions) --- from the original charge and from its image.

In order to illustrate some of these features, let us consider a pointlike charge (an electron) approaching from a vacuum the flat boundary
of a medium with a general $\varepsilon(\omega)$, Fig.~\ref{fig-geometry}, right.
We assume an oblique incidence with an angle $\alpha$ between the particle trajectory and the interface normal. We then define the incidence plane 
and the direction of the specular reflection. Any direction of the emitted photon can be characterized by the two ``flat'' angles $\theta_1$ and $\theta_2$ describing 
the out-of-the-plane deviation angle ($\theta_2$) and the in-the-plane projection measured from the specular reflection direction ($\theta_1$).

The classical result of Ginzburg and Frank concerns the normal incidence, for which only one polar angle $\theta$ is needed, 
which is measured from $-z$ for the backward TR geometry shown in our picture.
When solving TR problem for an oblique incidence, it is easier to work with the usual polar and azimuthal angles $\theta, \phi$, 
but the final result is more illustrative when expressed via the ``flat'' angles $\theta_{1,2}$. 
Below we shall use both pairs of the variables which are related as follows:
\be
\cos \theta = \cos \theta_2 \cos (\alpha + \theta_1),\quad \sin \phi = {\sin \theta_2 \over \sqrt{1 - \cos^2\theta_2 \cos^2(\alpha + \theta_1)}}\,.
\ee
These two angles can also be given another interpretation. Consider the spherical coordinate system with respect to the axes $(x',y',z') = (-z,x,y)$.
Then, the spherical angles $\theta'$ and $\phi'$ are nothing else but $\pi/2-\theta_2$ and $\alpha+\theta_1$, respectively.
Therefore, the measure for the angular integration is simply
\be
d\Omega = \sin\theta d\theta d\phi = \cos\theta_2 d\theta_2 d\theta_1\,.\label{measure}
\ee
The general formula for the charge TR spectral-angular distribution of the emitted energy for an oblique incidence 
on an ideally conducting ($\varepsilon^{\prime\prime} \rightarrow \infty$) target is \footnote{This formula as well as the others given below can be deduced from the expressions provided in Sec.~\ref{section-formulas}, and they coincide with the familiar results of Refs.~\cite{GTs, Paf, TM, G}.}
\be
{d^2 W \over d\omega d\Omega} = {e^2 \over \pi^2 c} \beta^2 \cos^2\alpha {{(\sin \theta - \beta \sin \alpha \cos \phi)^2 + \beta^2 \sin^2 \alpha \cos^2 \theta \sin^2 \phi}\over{((1-\beta \sin \alpha \sin \theta \cos \phi)^2 - \beta^2 \cos^2 \theta \cos^2 \alpha)^2}} \,,
\ee
both for the forward TR and for the backward TR.	

In the non-relativistic approximation and at the normal incidence, it has a typical form for any dipole radiation:
\be
{d^2 W \over d\omega d\Omega} \approx {e^2 \over \pi^2 c} \beta^2 \sin^2\theta\,.
\ee
As the electron becomes relativistic, the angular dependence develops two prominent lobes near the forward and backward directions with the maxima at $\theta = \gamma^{-1} \ll 1$:
\be
{d^2 W \over d\omega d\Omega} \approx  {e^2 \over \pi^2 c } {\theta^2 \over (\gamma^{-2} + \theta^2)^2} \,.\label{d2W-charge-UR}
\ee
At an oblique incidence, the two lobes shift. The forward TR is located near the particle trajectory, while the backward lobe stays close to the specular reflection direction. 
In the relativistic case, $\theta_1, \theta_2, \gamma^{-1} \ll 1$, the angular distributions become slightly asymmetrical in the incidence plane (direction quantified by $\theta_1$), 
but stay symmetric (in the absence of a magnetic moment) in the orthogonal plane:
\be
{d^2 W \over d\omega d\Omega} \approx {e^2 \over \pi^2 c } {\theta_1^2 + \theta_2^2 \over (\gamma^{-2} + \theta_1^2 + \theta_2^2)^2} 
{1 \over (1 - \theta_1 \tan \alpha)^2} \,.
\ee
The typical width of the lobes in the wave zone is $\sim \gamma^{-1}$. Note that for the moderately relativistic electrons, 
for example those produced in the electron microscopes
($E_e= 300$ keV, $\beta \approx 0.8$), the lobes are rather wide and are sizably shifted with respect to the reference directions. 

In the general case of a finite $\varepsilon (\omega)$, the radiation lobes in the backward/forward directions stay asymmetric in $\theta_1$ but do not coincide. 
It happens, in particular, for the almost transparent media due to a possible contribution of Cherenkov radiation in the forward direction
(we remind the reader that Cherenkov radiation and TR are two faces of fundamentally the same process of polarization radiation). 
If the target is a good conductor, the energies emitted in the forward/backward directions coincide. 
However for the medium with a weak absorption ($\varepsilon^{\prime \prime} \ll \varepsilon^{\prime}$), the interface reflectivity is small, 
so the forward TR dominates.

The TR photon spectrum is mostly shaped by the medium dispersion, $\varepsilon(\omega)$. For the forward TR (the energetic photons go in the forward direction only just due to the Doppler effect), the spectrum stays roughly flat below the critical frequency $\omega_c \sim \gamma \omega_p$. Above the plasma frequency, $\omega \gg \omega_{p}$, the medium becomes increasingly transparent with a typical dependence $\varepsilon - 1 \propto 1/\omega^2$, which leads to a rather sharp cut-off in the spectrum when $\omega \gg \omega_c$. 
This implies that for the moderately relativistic electrons, TR detection beyond the optical/UV spectral region is difficult.

Finally, the target can also be a highly conducting film,  sufficiently thin to let the incident charge cross both boundaries
without changing significantly its velocity, but at the same time thick enough (much thicker than the skin depth in the medium) 
to absorb any in-medium radiation.
In this case, both forward and backward TR will be observed, but they are emitted at different stages of the process:
the {\em detectable} backward TR is emitted in a vacuum when the charge enters the medium, 
while the {\em detectable} forward TR, again in a vacuum, is emitted when it exits the medium.
Although Fig.~\ref{fig-geometry} and the above discussion refer only to the former case, 
our detailed calculations below will include both cases.

To avoid any confusion, let us explicitly state list the kinematical conventions we use.
When presenting the results for the TR, we will always assume that it refers to TR in vacuum,
and in these circumstances, the distinction ``backward/forward TR'' should be understood as
the backward TR upon entering the medium and the forward TR upon exiting the medium.
This convention is natural as it matches the forward/backward radiation a photon detector in a typical experiment would observe.
In both cases, the normal ${\bm n}$ points to the hemisphere which the particle moves into, so that $({\bm u} {\bm n}) > 0$.
That is, at the first crossing, ${\bm n}$ points inside the medium, while at the second crossing it points outside, into the vacuum.
The coordinates $(x,y,z)$ are always the same as shown in Fig.~\ref{fig-geometry}, right; in particular, ${\bm n} = (0,0,1)$.
On the other hand, the angles $\theta$ and $\theta_1$ change in a corelated manner. The angle $\theta$ is always measured 
with respect to the normal pointing into the vacuum, that is, from $-z$ in the former case and from $z$ in the later case.
The angle $\theta_1$ is measured from the direction of specular reflection in the former case,
and from the actual trajectory of the charge upon its exit in the latter case. The angle $\theta_2$ is the same in both cases.

\subsection{TR from a magnetic moment}

TR from a pointlike neutral particle carrying a non-zero magnetic moment was considered, for instance, in \cite{GTs}.
Theoretical description of this process must address several delicate aspects.
The first subtlety is that the magnetic moment can be modelled, classically,
either as a close pair of magnetic monopoles or as a current loop of a small size.
It is remarkable that in a generic situation (arbitrary orientation of the magnetic moment
and an arbitrary permeability of the medium) these two approaches
lead to distinct results, both for the TR energy and for the polarization of the emitted radiation \cite{GTs}. 
A similar ambiguity appears for Cherenkov radiation, see e.g., \cite{Frank}.
Therefore, it should be stressed that, in the absence of magnetic monopoles in Nature,
we should always model the magnetic moment by a current loop.

The second subtlety is that the electric and magnetic dipole moments are not invariant
upon Lorentz boosts. In general, the electric and magnetic dipole moments transform as
the components of an antisymmetric tensor $M^{\mu\nu}$. 
If ${\bm \mu}$ is the magnetic moment in the particle rest frame (here and everywhere below, the bold face indicates
the 3D vectors),
then upon a boost with the velocity ${\bm u}$ it generates an electric dipole
moment ${\bf d} \| [{\bm u} \times {\bm \mu}]$. 
Fortunately, in the case of the vortex electron beams the magnetic moment is parallel to the average
propagation direction, which eliminates the electric dipole moment contribution.
The only effect then is the Lorentz contraction of the magnetic moment value
from $\mu$ in the rest frame to $\mu/\gamma$ in the lab frame. 
Since, as explained in the Introduction, we shall neglect all the quantum effects, the magnetic moment is not flipped during the emission.

The main changes of the TR from a longitudinal pointlike magnetic dipole $\mu = \ell \mu_B$
with respect to the charge TR can be anticipated already from comparison between the respective currents:  
${\bm j}_{\mu} = c\ \rot [{\bm \mu} \delta({\bm r} - {\bm u}t)]/\gamma$ vs. ${\bm j}_e = e\, {\bm u}\,\delta({\bm r} - {\bm u} t)$.
Curl leads to an extra factor $i\omega/c$ in the Fourier components of the radiation field.
As a result, the relative strength of the magnetic moment TR always bears the following small factor
\be
x_\ell = \ell {\hbar \omega \over E_e}\,.\label{xell}
\ee
The radiation energy contains this factor squared. For the optical/UV photons and for the typical electron energies achievable in an electron microscope, 
we get 
$$
x_\ell \sim 10^{-5}\ell.
$$ 
Therefore, radiation of the pure magnetic moments is suppressed by several orders of magnitude. 
Increasing of $\ell$ partially compensates this suppression, but it still remains prohibitively difficult to detect.

As we shall demonstrate below, the general formula for the ``pure'' magnetic moment TR for an oblique incidence on an ideally conducting target 
(again, identical for the backward TR and the forward TR) is
\begin{equation}
{d^2 W \over d\omega d\Omega}\Big|_{\mu} = \gamma^{-2}{\mu^2 \over \pi^2 c} \Big ({\omega \over c}\Big )^2 {\sin^2 \alpha \sin^2 \phi (1 - \beta \sin \alpha \sin \theta \cos \phi)^2 + \cos^2 \theta [\beta \sin \theta (1 - \sin^2 \alpha \sin^2 \phi) - \sin \alpha \cos \phi]^2 \over [(1-\beta \sin \alpha \sin \theta \cos \phi)^2 - \beta^2 \cos^2 \theta \cos^2 \alpha]^2} \,. \label{mu}
\end{equation}
Taken at face value, this expression does not vanish when $\beta \rightarrow 0$. 
However, as will become clear below, it does so for any finite $\varepsilon (\omega)$, which simply means that an ideal conductor 
as a model has limited applicability. 
When $\mu \approx \ell \mu_B$, we have
$$
\gamma^{-1}{\mu \omega \over c} = {1 \over 2} e x_\ell
$$

At the normal incidence and in the ultrarelativistic case, we have a formula which is very similar to (\ref{d2W-charge-UR}):
\be
{d^2 W \over d\omega d\Omega}\Big|_{\mu} \approx \gamma^{-2}{\mu^2 \over \pi^2 c }\Big ({\omega \over c}\Big )^2 {\theta^2 \over (\gamma^{-2} + \theta^2)^2} \,.
\ee
Finally, the relative intensity of the magnetic moment radiation, again at the normal incidence, is
\be
{d^2 W \over d\omega d\Omega}\Big|_{\mu}\bigg/{d^2 W \over d\omega d\Omega}\Big|_e = \Big ({ \gamma^{-1} \mu \omega \cos \theta \over e c}\Big )^2 \approx {1 \over 4} x^2_\ell \cos^2 \theta \ll 1
\ee
For an oblique incidence, the angular distributions of the magnetic moment TR are also asymmetric in the incidence plane (with respect to $\theta_1$), 
but stay symmetric in the perpendicular plane (in $\theta_2$).

\subsection{TR from charge + magnetic moment}

Of course, in the case of an electron, we deal with both the charge and the magnetic moment contributions
to TR. The fields of both sources add up, and the radiated energy can contain three terms
\be
dW = dW_e + dW_{e\mu} + dW_\mu\,,\label{dW3}
\ee
describing the radiation energy of the charge $dW_e$ and of the magnetic moment $dW_\mu$
as well as their interference $dW_{e\mu}$. The explicit equations for $dW$ will be given in the next Section.

If we want to detect TR from the magnetic moment in a situation with an extremely small $dW_\mu$,
we should focus on extracting the interference term $dW_{e\mu}$.
This task turns out to be tricky due to a number of reasons. An analysis of the situations when
this interference is present was performed in \cite{PKS}.

First, the emitted energy is a $3$-scalar while ${\bm \mu}$ is a pseudovector.
Therefore, the interference term must contain a triple product ${\bm e}_k\cdot [{\bm \mu\, \bm n}]$,
where ${\bm e}_k$ is the direction of the emitted photon, and ${\bm n}$ is the boundary normal.
This triple product vanishes for the normal incidence, while for an oblique incidence it changes sign
upon $\theta_2 \to -\theta_2$ (i. e. by flipping the sign of the out-of-the-plane component of ${\bm e}_k$).
Therefore, the interference can be observed only at an oblique incidence and only in the differential distribution,
not in the total energy.

Since the interference term is small compared to the pure charge radiation,
the angular distribution will also contain two lobes in the forward/backward direction,
but they will be slightly non-symmetric under $\theta_2 \to -\theta_2$.
A convenient way to quantify this distortion is to calculate the asymmetry
\be
A(\alpha, \omega, \ell) = \frac{\displaystyle \int d\Omega\, f(\theta_2) {d^2W \over d \omega d\Omega}}
{\displaystyle \int d\Omega\, |f(\theta_2)| {d^2W \over d \omega d\Omega}}\,.\label{asymmetry_0}
\ee
where $f(\theta_2)$ is some function, odd in $\theta_2 \rightarrow -\theta_2$. 
The simplest choice, $f(\theta_2) = \sgn (\theta_2)$, yields the widely used expression
\be
A = \frac{\displaystyle \int d\Omega_L\, {d^2W \over d \omega d\Omega} - \displaystyle \int d\Omega_R\, {d^2W \over d \omega d\Omega}}
{\displaystyle \displaystyle \int d\Omega_L\, {d^2W \over d \omega d\Omega} + \displaystyle \int d\Omega_R\, {d^2W \over d \omega d\Omega}}\,.\label{asymmetry}
\ee
Here, $d\Omega_L$ and $d\Omega_R$ indicate two hemispheres lying to the left and to the right from the incidence plane.
In fact, these integration domains do not have to cover the entire hemispheres, but in any case they must be symmetric under $\theta_2 \to -\theta_2$. 
Alternative definitions of the asymmetry, in which one weights the angular distribution, say,
with the function $f (\theta_2) = \sin \theta_2$, can also be employed. 
Below we shall use the definition (\ref{asymmetry}) unless explicitly mentioned otherwise.

There is yet another factor that can suppress interference.
Note that the curl, which is present in the definition of ${\bm j}_{\mu}$,
produces an extra $i$ factor in the Fourier components. As a result, the radiation field will contain the magnetic moment contribution with a relative phase:
\be
{\bm H}^R = {\bm H}_e^R + {\bm H}_\mu^R = a + i x_\ell b\,,
\ee
with some quantities $a$ and $b$. These two quantities are, generally speaking, complex 
due to the complex $\varepsilon$ (or, to be more accurate, due to the complex $\sqrt{\varepsilon}$).
However, if they have equal phases, the interference term $dW_{e\mu}$ vanishes.
This happens, in particular, in the two limiting cases: 
\begin{itemize}
\item
$\Im\, \varepsilon = 0$, a transparent medium;
\item
$\Im\, \varepsilon = \infty$, an ideal conductor. 
\end{itemize}
Therefore, in order to get a non-zero asymmetry, we must consider a real medium
with a sizable (but not asymptotically large) $\Im\, \varepsilon$.

If all these conditions are satisfied, we can expect, very roughly, 
the asymmetry (\ref{asymmetry}) of the order of $A \sim x_\ell$.
For typical experiments with vortex electrons in the microscopes,
this amounts to $A \sim {\cal O}(1\%)$ for the optical/UV TR from 
the electrons with $\ell \sim {\cal O}(1000)$, and the proportionally weaker
asymmetries for smaller $\ell$.

This makes detection of the asymmetry a rather delicate experimental undertaking.
It necessitates a careful numerical analysis of the effect, which we perform below.
It will allow us to obtain a reliable numerical results for the realistic setups and 
to check how this asymmetry can be enhanced.

We end this Section by mentioning that there exists alternative suggestion to detect 
the large OAM effect in transition radiation, \cite{PK}, which relies on the recent
calculations \cite{PKS}. In this method, the quantity of interest is not the angular distribution 
of the emitted photons but their polarization. Without the magnetic moment contribution,
the emitted photons are linearly polarized.
The presence of the magnetic moment leads to a slightly elliptical polarization
for the off-plane photons. If one manages to measure the photons polarization
very close to the direction of the minimum intensity, the degree of circular
polarization can be sizable, of the level of few percent or higher for $\ell = 100$.
Whether such an accurate angular selection is feasible in realistic devices, remains to be studied.

\section{TR from vortex electrons: quantitative description}\label{section-formulas}

\subsection{Vortex electrons}

A vortex electron state is a freely propagating electron whose wave function
contains phase singularities with a nonzero winding number $\ell$.
Such an electron state is characterized, simultaneously, by an average propagation direction
and an intrinsic orbital angular momentum (OAM) with the projection $L = \hbar \ell$ on this direction. 
Following the suggestion of Ref.~\cite{Bliokh-07}, vortex electrons were recently created in experiments by several groups, 
\cite{vortex-experimental-tonomura,vortex-experimental-verbeeck,vortex-experimental-mcmorran,vortex-experimental-uchida}.
They are produced in electron microscopes with the typical energy of $E_e = 200-300$ keV with the aid of 
the computer generated diffraction gratings, which induce $\ell$ as large as $25$ in the first diffraction peak
and proportionally larger $\ell$ in faint higher diffraction peaks.
These vortex electrons can be accurately manipulated and, in particular, can be focused to a spot of an angstrom size
\cite{verbeeck-2}.

The simplest example of a vortex state for a spinless particle is given by the Bessel beam state 
\cite{serbo,K-2012}
whose coordinate wave function is
\be
\psi(r_\perp, \phi_r, z) \propto e^{ik_z z} e^{i\ell \phi_r} J_{\ell}(k_\perp r_\perp)\,.
\ee
At large $\ell$, the properties of the Bessel functions lead to a narrow radial distribution 
located around $r_\perp \approx \ell /k_\perp$, in a good analogy with 
the quasiclassical picture of such an electron as a rotating ring of electronic density.

The spin degree of freedom of the vortex electron can also be included \cite{Bliokh-11,K-2012}.
Spin and OAM degrees of freedom interact \cite{Bliokh-11}, and both of them
induce the magnetic moment of the vortex electron (in the lab frame)
\be
{\mu \over \gamma} = (\ell + 2s - \Delta s){\mu_B \over \gamma} \approx \ell{\mu_B \over \gamma}\label{mu-ell}\,,
\ee
which was confirmed by the observation of the OAM-dependent
Larmor precession in the longitudinal magnetic field, \cite{larmor-precession}.
Here $\Delta s$ is an effective shift in the magnetic moment due to spin-orbital
interaction. 
In the case of large $\ell$, which concerns us in this paper, we can neglect
the spin contribution, which is indicated in the last expression in (\ref{mu-ell}).
The OAM-induced magnetic moment in this approximation
is aligned with the average propagation direction of the vortex electron regardless of the spin state.

\subsection{Modelling large OAM-induced magnetic moment}

An electron vortex state is characterized by a nontrivial spatial structure of 
the wave function. In this sense, it is an inherently quantum state.
However, as we explained in the Introduction, the large value of $\ell$
allows one to treat PR from the OAM-induced magnetic moment
quasiclassically neglecting the quantum effects during radiation, because 
the latter is of order $\hbar \omega/E_e$, which is much less than the OAM contribution, $\ell \hbar \omega/E_e$.
 
Not only does the magnetic moment (\ref{mu-ell}) describe how vortex electrons 
couple to an external magnetic field,
but it is also a source of its own EM field.
Therefore, if the vortex electron wavepacket is sufficiently compact,
it can be modelled as a classical {\em pointlike} source with a charge $e$ and
an {\em intrinsic} magnetic moment $\mu$ given by (\ref{mu-ell}), Fig.~\ref{fig-geometry}.
This picture is behind our first method of calculating TR from
the vortex electrons passing from one medium into another.
In this purely phenomenological model, we do not discern the internal microscopic structure
of the vortex electron, nor do we specify the origin of the large magnetic moment.
The only assumption we make is that, in the absence of magnetic monopoles,
the magnetic moment arises only from closed charge current loops.

\begin{figure}[!htb]
   \centering
\includegraphics[width=6cm]{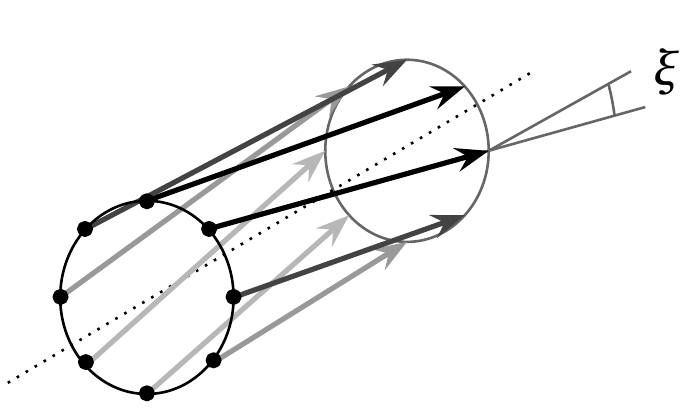}
\caption{Modelling magnetic moment of a vortex electron via a flat thin rotating ring 
of point charges.}
   \label{fig-geometry2}
\end{figure}

To control the validity of this approach, we devised our second model,
which also treats the vortex electron quasiclassically but in which
the OAM-induced magnetic moment becomes an {\em emergent} quantity.

In this model, we calculate {\em coherent} transition radiation from a 
charged rotating ring consisting of a large number of electrons, $N \gg 1$, 
which carry no intrinsic magnetic moment and whose trajectories are straight rays passing at fixed skew angles through a ring 
of a microscopic size, $R \ll \lambda$, see Fig.~\ref{fig-geometry2}.
Individual charges move at constant and equal longitudinal velocities, so that at any given moment of time they form an infinitely thin annular slab
in the transverse plane. It then becomes the standard calculation of TR 
with the only exception that the total charge of the ring is just $e$
(in other words, we calculate the coherent radiation energy of a ring 
and divide it by the factor $N^2$). Note that this thin ring model is qualitatively similar
to the true transverse wave function profile of a large-$\ell$ vortex electron
mentioned above.

In order to compare the two models, we need to determine the effective $\ell$ 
within the second approach.
This can be done quasiclassically as follows:
\be
\ell_{\rm eff} = {R p \sin\xi \over \hbar}\,,
\ee
where $p$ is the electron momentum and $\xi$ is the skew angle, so that $p \sin\xi$ is the absolute
value of the transverse momentum of each electron in the ring.
The same expression can be also obtained
from the definition of the magnetic moment of a current loop with an area $S$ and the current $I$ 
\be
\ell_{\rm eff} = {\mu\over \mu_B} = {2m_e c \over e\hbar } \gamma \pi R^2 {e v \sin\xi \over 2\pi R c}
= {R p \sin\xi \over \hbar}\,.
\ee
As usual, $\mu$ refers to the magnetic moment in the rest frame, while the expression $S I/c$
gives the magnetic moment in the lab frame. 

These models can be applicable to a realistic experimental set-up with vortex electrons,
if certain coherence conditions are satisfied. 
First, the quasiclassical treatment of the electrons as pointlike particles
in the transverse space is valid only if the vortex electrons are focused in a 
spot with a size much smaller than the emitted light wavelength $\lambda$.
Within the second model, we also assume that the size of the ring is smaller
than $\lambda$ in order to avoid destructive interference between different parts of the electron wave function.
The same applicability condition requires also that the longitudinal extent of the 
individual electron wave function is much shorter than $\lambda$.
This extent can be quantified by the longitudinal self-correlation length of the electron beam.
This length is related to the monochromaticity of the electron beam and
it can be found experimentally by counting the number of fringes
in an electron diffraction experiment.
The longitudinal compactness condition implies that the monochromaticity
should not be too good.

Finally, the calculations of TR presented below are performed for individual electrons, not electron bunches
because we assume that successive electrons pass through the foil one at a time.
This condition is, in fact, an important part of the whole idea of making use of vortex electrons.
A vortex electron state refers to a state of a single sufficiently isolated electron, whose wave function remains stable over long distances
due to the absence of disturbance of copropagating electrons. It is hard to imagine that a compact dense electron
bunch would be able to keep each electron in a definite vortex state.
This condition implies that electrons must be separated by distances much larger than $\lambda$,
which in turn restricts the current to values below $\sim 10 \mu$A.
This is satisfied by a large margin in experiments with vortex beams realized so far.

If the above coherence requirements are all fulfilled, 
the two models are expected to yield qualitatively similar and numerically close results,
because the second model proposes a microscopic origin of 
the large magnetic moment introduced ``by hand'' 
in the first model. We notice that for the simple case of the normal incidence and a pure magnetic moment (no charge),
a similar expectation was explicitly mentioned and verified in chapter $3.7$ of \cite{GTs}.

\subsection{Methodical example: Cherenkov radiation from charge + intrinsic magnetic moment}

We start with the simpler case of Cherenkov radiation by vortex electrons with large
OAM-induced magnetic moments, which is calculated according to the first model.

We consider a pointlike particle with a charge $e$ and an intrinsic magnetic moment, 
which in the particle rest frame is equal to ${\bm \mu}$ and directed along the velocity. 
The charge and current densities in the rest frame in vacuum are
\begin{eqnarray}
\displaystyle \rho_e = e \delta({\bm r}),\ {\bm j}_{\mu} = c\ \rot [{\bm \mu} \delta({\bm r})]\,. \label{Eq1}
\end{eqnarray}
Note that this expression is valid for the case when the magnetic moment originates from current loops.

In the lab frame, the currents are
\begin{eqnarray}
{\bm j}_e = e\, {\bm u}\,\delta({\bm r} - {\bm u} t),\quad 
{\bm j}_{\mu} = {c \mu \over \gamma}
\left(\begin{array}{c}
\partial_y \\
- \partial_x \\
0
\end{array}\right)
 \delta({\bm r} - {\bm u} t)\,. \label{Eq2Ch}
\end{eqnarray}
Note that the Lorentz transformation induced decrease of the magnetic moment in the lab frame.
Their Fourier transforms\footnote{We use the integration measures $d^3 x/(2\pi)^3$ and $d^3 q$.} are
\be
\displaystyle {\bm j}_e ({\bm q}, \omega) = \frac{e}{(2\pi)^3} {\bm u}\, \delta(\omega - {\bm q}\cdot{\bm u}),\quad 
{\bm j}_{\mu}({\bm q}, \omega) = \frac{ic}{(2\pi)^3}\, {\bm e}_{\mu}\,\delta(\omega - {\bm q}\cdot{\bm u})\,,
\label{currents-method1}
\ee
where
\be
{\bm e}_{\mu} = {\mu \over \gamma} 
\left(\begin{array}{c}
q_y \\
- q_x \\
0
\end{array}\right)\,. \label{emu-method1Ch}
\ee
Note the all-important $i$ factor in the magnetic moment contribution. 

These currents generate electric fields which are determined by the Maxwell equations.
Generally, their Fourier components are
\begin{eqnarray}
\displaystyle {\bm E} ({\bm q}, \omega) = \frac{4\pi i}{\omega} \frac{1}{{\bm q}^2 - \omega^2/c^2} 
\left[\left(\frac{\omega}{c}\right)^2 {\bm j} ({\bm q}, \omega) - {\bm q}\ ({\bm q} \cdot {\bm j} ({\bm q}, \omega))\right]\,.\label{Eqomega-gen}
\end{eqnarray}
According to the polarization currents approach developed in Ref.~\cite{JETP}, the radiation field in the wave zone is found as
\begin{eqnarray}
{\bm H}^{R} ({\bm r}, \omega) = (2 \pi)^3 \left(\omega \over c\right)^2 {\varepsilon - 1 \over 4 \pi}\frac{e^{i\sqrt{\varepsilon}r\omega/c}}{r} {\bm e}_{k} \times \left ({\bm E}_e({\bm k}, \omega) + {\bm E}_{\mu}({\bm k}, \omega)\right) = \\ = i \frac{\omega^2}{c^3} \left(\varepsilon - 1\right)\,\frac{e^{i\sqrt{\varepsilon}r\omega/c}}{r}\, \frac{\delta (\omega - {\bm k} \cdot {\bm u})}{{\bm k}^2 - \omega^2/c^2}{\bm e}_{k} \times \left[e \bm{u} + i c {\mu \over \gamma} {\bm k} \times {\bm e}_u \right]\,,\label{HRCh}
\end{eqnarray}
and the argument of the delta-function turns into zero under the Cherenkov condition, $1 = \beta \sqrt{\varepsilon} \cos \theta_m$. 
Here, ${\bm e}_u = {\bm u}/u = (0, 0, 1)$ and ${\bm k} = \omega {\bm e}_k /c = \omega \sqrt{\varepsilon} (\sin \theta_m \cos \phi, \sin \theta_m \sin \phi, \cos \theta_m)/c$ 
is the wave vector. Calculating the radiated energy as
\be
{d^2 W \over d\omega d\Omega} = {c r^2 \over \sqrt{\varepsilon}} |{\bm H}^R|^2,
\ee
we evaluate the squared delta-function in the usual way,
$$
\delta^2 (\omega - {\bm k} \cdot {\bm u}) \rightarrow \delta (\omega - {\bm k} \cdot {\bm u}) \delta(0) \rightarrow {T \over 2\pi} \delta (\omega - {\bm k} \cdot {\bm u}),
$$
where $T$ is a large ($T \gg \omega^{-1}$) period of time. Integrating the resultant expression over the angles we note that the delta-function zero lies on the integration path only if the permittivity $\varepsilon$ is real ($\varepsilon^{\prime \prime} = 0$). In that case we come finally to:
\be
{1 \over u T}{d W \over d\omega} = {e^2 \over c^2} \omega \left(1 - {1 \over \beta^2 \varepsilon}\right) \left[1 + \left({\mu \omega \sqrt{\varepsilon} \over e u \gamma}\right)^2 \right].
\ee
This is the Tamm-Frank formula for Cherenkov radiation with a contribution of the magnetic moment. As predicted, there is no interference term, $dW_{e\mu}$, due to transparence of the medium being considered. 

It should be noted, however, that this term is absent even in an absorbing medium. Indeed, for a medium with a weak absorption (otherwise, the Cherenkov radiation problem itself has no sense whatsoever in a boundless medium), the radiation field squared is proportional to (here $\kappa \equiv \Im \sqrt{\varepsilon}$)
\be
\Big |{\bm e}_{k} \times [e \bm{u} + i c \gamma^{-1}\mu\,{\bm k} \times {\bm e}_u ]\Big |^2 \propto (e u \sin \phi - \kappa \gamma^{-1} \mu \omega \cos \theta \cos \phi )^2 + (e u \cos \phi + \kappa \gamma^{-1} \mu \omega \cos \theta \sin \phi )^2
\ee
and the terms linear in $\mu$ cancel each other. This remarkable feature is obviously due to the azimuthal symmetry of the problem. 
This is not the case for transition radiation in the oblique incidence geometry, which we are now going to demonstrate.

\subsection{Radiation field for TR from charge + intrinsic magnetic moment}\label{method_1}

Now we consider TR generated by an oblique passage of a particle with a charge and a magnetic moment through a flat interface between a vacuum and 
a non-magnetic medium with a (complex) permittivity $\varepsilon (\omega)$. 
Axis $z$ is chosen as the normal to the interface, and axis $x$ defines the particle incidence plane.
The particle approaches the boundary in the $(x,z)$ plane at the angle $\alpha$ to the normal, 
and its velocity is ${\bm u} = u (\sin \alpha, 0, \cos \alpha)$.

In the lab frame, the currents are
\begin{eqnarray}
{\bm j}_e = e\, {\bm u}\,\delta({\bm r} - {\bm u} t),\quad 
{\bm j}_{\mu} = {c \mu \over \gamma}
\left(\begin{array}{c}
\cos \alpha\ \partial_y \\
\sin \alpha\ \partial_z - \cos \alpha\ \partial_x \\
- \sin \alpha\ \partial_y
\end{array}\right)
 \delta({\bm r} - {\bm u} t)\,. \label{Eq2}
\end{eqnarray}
Their Fourier transforms stay the same, (\ref{currents-method1}), with
\be
{\bm e}_{\mu} = {\mu \over \gamma} 
\left(\begin{array}{c}
\cos \alpha\ q_y \\
\sin \alpha\ q_z - \cos \alpha\ q_x \\
- \sin \alpha\ q_y
\end{array}\right)\,. \label{emu-method1}
\ee
In the problem of calculating TR, we deal with a situation which is homogeneous along the coordinates $x$ and $y$, but not along $z$
due to the presence of a boundary. Therefore, it is convenient to work with the partial Fourier transforms, 
${\bm E}({\bm q}_{\perp}, z, \omega)$, with ${\bm q}_{\perp} = (q_x, q_y, 0)$
in which the dependence on $z$ is kept.
Due to linearity, the electric field (\ref{Eqomega-gen}) is a sum of the contributions 
from both currents (\ref{currents-method1}), which can be written as follows:
\begin{eqnarray}
{\bm E}_e ({\bm q}_{\perp}, z, \omega) &=& i \frac{2 e}{(2 \pi)^2\omega\ u_z}\,
\frac{e^{iz (\omega - {\bm q}_{\perp}\!\cdot{\bm u})/u_z}}{{\bm q}_{\perp}^2 + (\omega - {\bm q}_{\perp}\cdot{\bm u})^2/u_z^2 - \omega^2/c^2} 
\left[\left(\frac{\omega}{c}\right)^2 {\bm u} - \omega \left({\bm q}_{\perp} + {\bm n} \frac{\omega - {\bm q}_{\perp}\!\cdot{\bm u}}{u_z}\right)\right],\\
{\bm E}_{\mu} ({\bm q}_{\perp}, z, \omega) &=& -\frac{2 e}{(2 \pi)^2\omega\ u_z}
\frac{e^{iz (\omega - {\bm q}_{\perp}\!\cdot{\bm u})/u_z}}{{\bm q}_{\perp}^2 + (\omega - {\bm q}_{\perp}\cdot{\bm u})^2/u_z^2 - \omega^2/c^2} 
\times \nonumber\\
&&\qquad \qquad\times
\left[\left(\frac{\omega}{c}\right)^2 {\bm e}_{\mu} - \left({\bm q}_{\perp} + {\bm n} \frac{\omega - {\bm q}_{\perp}\!\cdot{\bm u}}{u_z}\right) 
\left({\bm q}_{\perp}\!\cdot {\bm e}_{\mu} + \frac{\omega - {\bm q}_{\perp}\!\cdot{\bm u}}{u_z} e_{\mu, z}\right)\right],\label{Eq5}
\end{eqnarray}
where ${\bm e}_{\mu}$ is given by (\ref{emu-method1}).

In order to calculate the TR field in the wave zone, we use the same polarization current technique.
The radiation field can be written as
\be
{\bm H}^{R} ({\bm r}, \omega) = \left(\frac{2\pi\omega}{c}\right)^2 \frac{\varepsilon - 1}{4 \pi} 
\,\frac{e^{i\sqrt{\varepsilon}r\omega/c}}{r}\, \left[{\bm e}_{k} \times \bm{\mathcal{J}}\right]\,,\label{HR}
\ee
where
\be
\bm{\mathcal{J}} = \int dz^{\prime} e^{-i z^{\prime} k_z} 
\left[{\bm E}_e ({\bm k}_{\perp}, z^{\prime}, \omega) + {\bm E}_{\mu} ({\bm k}_{\perp}, z^{\prime}, \omega)\right],\label{pol-current}
\ee
is a quantity proportional to the polarization current \cite{JETP}.
We introduced here the ``on-shell'' wave vector in the medium ${\bm k} = {\bm e}_k \omega/c$, where
\be
{\bm e}_k = \sqrt{\varepsilon} 
\left(\begin{array}{c}
\sin \theta_m \cos \phi \\
\sin \theta_m \sin \phi \\
\cos \theta_m
\end{array}\right)
= 
\left(\begin{array}{c}
\sin \theta \cos \phi \\
\sin \theta \sin \phi \\
\pm \sqrt{\varepsilon - \sin^2 \theta}
\end{array}\right)\,.\label{ek}
\ee
The two expressions in (\ref{ek}) relate the emission angle in the medium $\theta_m$ with the emission angle $\theta$ in a vacuum:
$\sqrt{\varepsilon}\sin\theta_m = \sin\theta$. 
Integration in (\ref{pol-current}) is carried out from $0$ to $\infty$ for the backward TR when the electron enters the medium
and from $-\infty$ to $0$ for the forward TR when it exist the medium.

It is instructive to stop for a moment and discuss the physical meaning of the quantities we manipulate with.
We work out the TR problem by applying the polarization current approach developed in detail in \cite{JETP}. 
In this approach we take the current itself as if the medium were boundaryless, 
$$
{\bm j} = \sigma ({\bm E}_e + {\bm E}_{\mu})\,,
$$
with $\sigma$ being a complex conductivity. 
Besides, the Green function pole is shifted, $\omega/c \rightarrow \sqrt{\varepsilon}\omega/c$, 
because of the effective ``dressing'' of the particle field in the medium (see e.g., \cite{Ryaz_QED}). 
The effects of the interface (or the interfaces) are taken into account when we find how this (bare) current field, 
which is calculated by integrating the current over the target volume, changes due to reflections and refractions at them. 
By applying the reciprocity theorem, we reduce the initial (rather complicated) problem to the \textit{complementary} problem of refraction, 
which is much easier to solve using the usual Fresnel laws and summing up all the secondary re-reflected fields inside the target. 
The necessity of using the reciprocity theorem may be argued, in fact, by the causality considerations, 
which require permittivity $\varepsilon (\omega)$ to be always a complex quantity; see also \cite{Bar}.

It is therefore not surprising that the quantities like the emitted photon ``direction'' ${\bm e}_k$ and its ``polar angle'' $\theta_m$
are complex. They correspond to a wave which is exponentially attenuated with propagation distance due to 
absorption by the medium, as is explicitly indicated by $\exp(i\sqrt{\varepsilon}r\omega/c)$ in (\ref{HR});
note that this defines the sign choice for $\sqrt{\varepsilon}$: $\Im  \sqrt{\varepsilon} > 0$.
Thus, we can formally manipulate with these quantities in the same way as we did for transparent media,
where they have a clear physical meaning. In this way, we can obtain expressions for the energy of the emitted radiation and its angular distribution, 
which are initially expressed in terms of complex ${\bm e}_k$ and $\theta_m$. However, we can then use the relation between $\theta_m$ and the true
polar angle for the radiation emitted in a vacuum $\theta$, and focusing on this case express the results in terms of $\theta$.
In this way, the complexity will be transferred from $\theta_m$ to $\sqrt{\varepsilon}$ or to the combination
\be
\sqrt{\varepsilon_\theta} \equiv \sqrt{\varepsilon - \sin^2\theta}\,, \nonumber
\ee
and the results will directly correspond to the radiation in a vacuum.

Continuing with the calculations,
the radiation field can be conveniently written in the coordinates related not with the electron incidence plane, 
but with the photon production plane $({\bm e}_k, z)$. 
The radiation field (\ref{HR}) is orthogonal to ${\bm e}_k$ and therefore has two components which lie in the production plane, $H^R_{in}$,
and out of that plane, $H^R_{out}$. In the vacuum variables, they are expressed as
\be
H^R_{out} = H^R_y \cos \phi - H^R_x \sin \phi\,,
\quad
H^R_{in} = {1 \over \sqrt{\varepsilon}} \left[-H^R_z \sin\theta \pm (H^R_x \cos\phi + H^R_y \sin\phi) \sqrt{\varepsilon_\theta}\right]\,.
\ee

The final expressions for these two components are
\begin{eqnarray}
H^R_{out} &=& {\cal N}\Big[ \sin \theta (1 - \beta^2 \cos^2\alpha - {\bm \beta} \cdot {\bm e}_k) \pm 
\beta^2 \sin \alpha \cos \alpha \cos \phi \sqrt{\varepsilon_\theta}  \nonumber\\
&& \quad +\ i \mu \frac{\omega}{e\gamma c} \sin \alpha \sin \phi \Big 
(\beta \cos \alpha \sin^2 \theta \mp \beta \sin \alpha \sin \theta \cos \phi \sqrt{\varepsilon_\theta} \pm \sqrt{\varepsilon_\theta}\Big )\Big]\,,
\label{Hout}\\
H^R_{in} &=& {\cal N}\sqrt{\varepsilon}\left[ \beta^2 \sin \alpha \cos \alpha \sin \phi 
+ i \mu \frac{\omega}{e\gamma c} \left[\beta \sin \theta (1 - \sin^2 \alpha \sin^2 \phi) - \sin \alpha \cos \phi\right]\right] \label{Hin}\,,
\end{eqnarray}
where the overall factor in front of the brackets is
\bea
{\cal N} &=& \pm \frac{e}{2 \pi c} (\varepsilon - 1) \beta \cos \alpha \frac{e^{i\sqrt{\varepsilon}r\omega/c}}{r} \times
\nonumber\\ && \qquad 
\times \Big [(1 - \beta \sin \alpha \sin \theta \cos \phi )^2 - (\beta \cos \alpha \cos \theta)^2\Big ]^{-1} \Big [1 - \beta \sin \alpha \sin \theta \cos \phi \mp \beta \cos \alpha \sqrt{\varepsilon_\theta}\Big ]^{-1}\,.
\eea
As before, the upper and lower signs in these expressions correspond to the forward radiation (upon exiting the medium)
and the backward radiation (upon entering the medium), respectively. 
As can be seen from the last expression, the radiation intensity vanishes in the limiting case $\beta \rightarrow 0$, 
both for the charge radiation and for the magnetic moment one.

The spectral-angular distributions of the radiated energy can be found from the reciprocity theorem as follows \cite{JETP}:
\begin{eqnarray}
&& \displaystyle \frac{d^2 W}{d\omega d\Omega} = 4 c r^2 \cos^2 \theta 
\left(\Big |\frac{1}{\varepsilon \cos \theta + \sqrt{\varepsilon_\theta}}\Big |^2 |H^R_{out}|^2 
+ \Big |\frac{1}{\sqrt{\varepsilon} (\cos \theta + \sqrt{\varepsilon_\theta})}\Big |^2 |H^R_{in}|^2 \right)\,.
\label{power-method1}
\end{eqnarray}
Substituting here the explicit expressions for the radiation field (\ref{Hout}) and (\ref{Hin}) 
and sorting out the charge and magnetic moment contributions, 
one can break the energy into the pure charge $dW_e$ and the pure magnetic moment $dW_\mu$ contributions 
as well as the interference term $dW_{e\mu}$, (\ref{dW3}).

It can be easily checked that for a neutral particle with a magnetic moment only and for an ideally conducting surface,
the resultant formula coincides with Eq.~(\ref{mu}).

We are interested in detecting a small contribution of the magnetic moment to the radiation energy.
There is a number of features which are visible directly in the above equations.
Firstly, with the value of the intrinsic magnetic moment (\ref{mu-ell}) and neglecting the spin contribution,
one sees that the interference term is indeed suppressed by the factor $x_\ell \ll 1$, 
while the pure magnetic moment contribution is $\propto x_\ell^2$.
For optical/UV photons and for the typical electron energies achievable in an electron microscope, we get $x_\ell \sim 10^{-5}\ell$.
This estimate makes it clear that one can only hope to detect the interference term $dW_{e\mu}$.
Secondly, it is plain to see that this interference term can originate only from $|H^R_{out}|^2$ and only with a nontrivially complex $\varepsilon$.
In particular, this interference term is absent for a transparent medium, $\Im\, \varepsilon = 0$ (similarly to the Cherenkov radiation case) 
and for the ideal conductor, $\Im\, \varepsilon = \infty$. 
Finally, this term also vanishes for the normal incidence ($\alpha = 0$) at any emission angles as well as at an oblique incidence 
for emission in the incidence plane, $\phi = 0$.

It should be noted that when considering TR at a grazing incidence of not very energetic electrons, the applicability conditions 
of macroscopic electrodynamics may be violated \cite{Ryaz_83}. So, the region where the models being used work well is determined by the following inequality
$$
{u_z \over \omega - ({\bm k}_{\perp}{\bm u})} = {\lambda \over 2\pi} {\beta \cos \alpha \over 1 - \beta \sin \alpha \cos \theta_2 \sin (\alpha + \theta_1)} \gg b\,,
$$
where $b$ is the interatomic distance ($\sim 1$ \AA). 
For the optical/near-UV region and the parameters considered below, the left-hand side of this inequality is ${\cal O} (0.1 \lambda \sim 10 \text{nm})$. 
However this condition may be violated for non-relativistic electrons ($E_e$ lower than $100$ keV) and/or for the angles of incidence $\alpha \rightarrow 90^{\circ}$.

\subsection{Radiation field from a charged ring with azimuthal current}\label{method_2}

Consider a particle moving at the angle $\xi$ to the $z$-axis. If its position in the plane $z = 0$ is given by the vector ${\bm \rho} = \rho (\sin \varphi, \cos \varphi, 0)$,
then its velocity is
$$
{\bm u} = u (- \sin \xi \cos \varphi,\, \sin \xi \sin \varphi,\, \cos \xi)\,.
$$
Note that $\varphi$ characterizes the particle position, while $\phi$ is still used to denote the azimuthal angle of the emitted photon.
The current thus acquires the azimuthal component, see Fig.~\ref{fig-geometry2}. 

For an oblique incidence on a screen, these expressions turn into
\be
{\bm \rho} \rightarrow \rho\, {\bm e}_{\rho} = 
\rho\, \left(\begin{array}{c}
\sin \varphi \cos \alpha \\
\cos \varphi \\
- \sin \varphi \sin \alpha
\end{array}\right)\,,
\quad
{\bm u} \rightarrow u\, 
\left(\begin{array}{c}
\cos \xi \sin \alpha - \sin \xi \cos \varphi \cos \alpha\\
 \sin \xi \sin \varphi \\
 \cos \xi \cos \alpha + \sin \xi \cos \varphi \sin \alpha
\end{array}\right)\,.
\label{Eq9}
\ee
Here $\alpha$ is the angle between the symmetry axis of the helical motion and the normal to the interface.

Since we assume that the charged particles following these trajectories have no intrinsic magnetic moment, TR is calculated in the standard way.
We only note that the Fourier transform of the current density for a trajectory with a given ${\bm \rho}$ acquires a phase factor 
\begin{eqnarray}
\displaystyle {\bm j}_e ({\bm q}, \omega) = \frac{e}{(2\pi)^3} \, {\bm u} \, \delta(\omega - {\bm q}\cdot{\bm u})\, e^{-i {\bm q} \cdot {\bm \rho}}\,. \label{Eq10}
\end{eqnarray}
We then find the radiation field for each ${\bm \rho}$ and integrate it over all polar angles $\varphi$ as well as with respect to $\rho$
within certain limits. It effectively corresponds to summing over a large number of particles distributed homogeneously
over a certain annular region. One can then recycle the formulas from the previous subsection by setting $\mu = 0$ there, 
and represent the out-of-the-plane and in-the-plane components for the radiation field as follows:
\begin{eqnarray}
H^R_{out} &=& \int_0^{2\pi} {d\varphi \over \pi} \int_{R_{min}}^{R_{max}} \!\!\!\!{\rho d\rho \over R_{max}^2-R_{min}^2}\, {\cal N}
\left[-\sin \theta \frac{\omega}{u_z} ({\bm k}\cdot {\bm u} - \omega (1 - \beta_z^2)) \pm \sqrt{\varepsilon_\theta} 
\Big (\frac{\omega}{c}\Big )^2 (u_y \sin \phi + u_x \cos \phi)\right]\,,\\
H^R_{in} &=& \int_0^{2\pi} {d\varphi\over \pi} \int_{R_{min}}^{R_{max}} \!\!\!\!{\rho d\rho \over R_{max}^2-R_{min}^2} \, {\cal N}\, \sqrt{\varepsilon}
\,\Big (\frac{\omega}{c}\Big )^2\, (u_x \sin \phi - u_y \cos \phi)\,,
\eea
where
\begin{eqnarray}
{\cal N} &=& \pm \frac{e\omega}{2\pi c^2} (\varepsilon - 1) \frac{e^{i\sqrt{\varepsilon}r\omega/c}}{r}	
{e^{-i \rho({\bm k}_{\perp}\! \cdot {\bm e}_{\rho} + e_{\rho, z} (\omega - {\bm k}_{\perp}\! \cdot {\bm u})/u_z)} \over
(\omega - {\bm k}\cdot{\bm u}) ({\bm k}_{\perp}^2 + (\omega - {\bm k}_{\perp}\! \cdot {\bm u})^2/u_z^2 - \omega^2/c^2)}
\eea
Integration over $\rho$ is trivial here, while integration over $\varphi$ may be performed numerically. The radiated energy is also found as in the previous case.

\section{Numerical results}\label{section-numerical}

\subsection{Benchmark case}

We start by presenting numerical results with the following choice of parameters, which we call the benchmark case. 
The medium is chosen to be aluminium (the permittivity data were taken from \cite{Rakic}), the incidence angle is $\alpha = 70^\circ$,
the electron energy is $E_e= 300$ keV. The TR lobes are broad functions of $\theta_1$ and $\theta_2$,
and are shown in Fig.~\ref{fig-bench3D}, where we plot the spectral-angular distribution at $\ell = 0$ as functions of $\theta_1$ and $\theta_2$. 

If we are aiming at detection of an asymmetry in $\theta_2$, we should focus on such a $\theta_1$ region in which
the $\theta_2$-dependence has a two-bump structure. 
For this purpose, we consider below $\theta_2$-distributions integrated over a $\theta_1$-region 
centered at some value $\bar\theta_1$; specifically, we choose the integration region $[\bar\theta_1 - 10^{\circ}, \bar\theta_1 + 10^{\circ}]$. 	
Then, at a non-zero and large $\ell$, we expect these two maxima to differ from each other.
In Fig.~\ref{fig-bench} we show the spectral-angular distribution of the emitted energy for the forward
and backward TR as a function of $\theta_2$ for the fixed $\bar\theta_1 = -40^\circ$ and $\hbar\omega = 5$ eV.
These choices constitute our benchmark case.

\begin{figure}[!htb]
   \centering
\includegraphics[width=8cm]{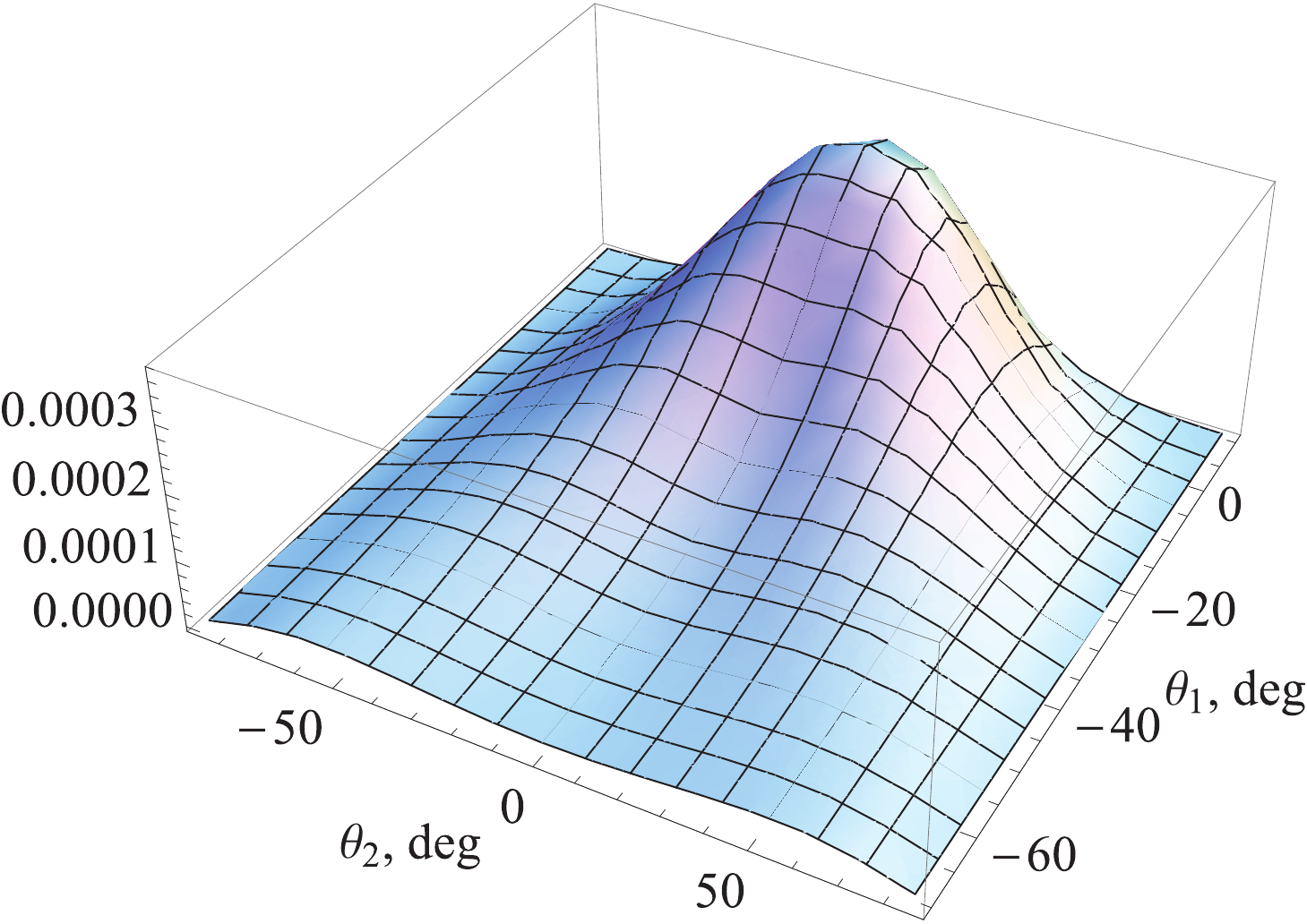}
\put(-240,110){ $\dst \fr{d^2W}{d\hbar\omega\, d\Omega}$}
\caption{(Color online.) Distribution of the emitted TR energy over the angles $\theta_1$ and $\theta_2$ in the benchmark case ($\alpha = 70^{\circ}, \gamma = 1.59, \hbar\omega = 5$ eV) at $\ell = 0$. The first model (see Sec.~\ref{method_1}) is used.}
   \label{fig-bench3D}
\end{figure}

\begin{figure}[!htb]
   \centering
\includegraphics[width=8cm]{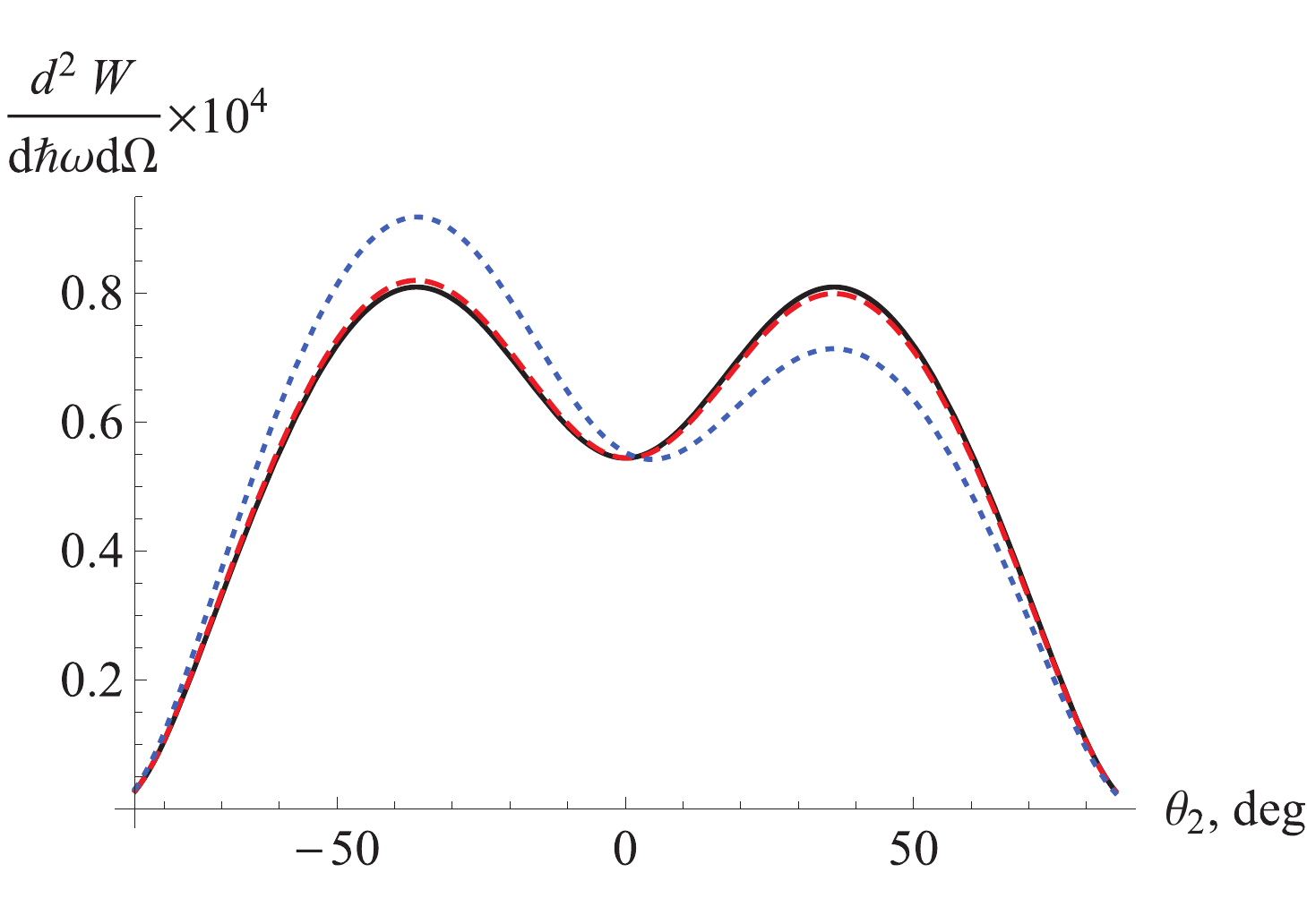}\hspace{5mm}
\includegraphics[width=8cm]{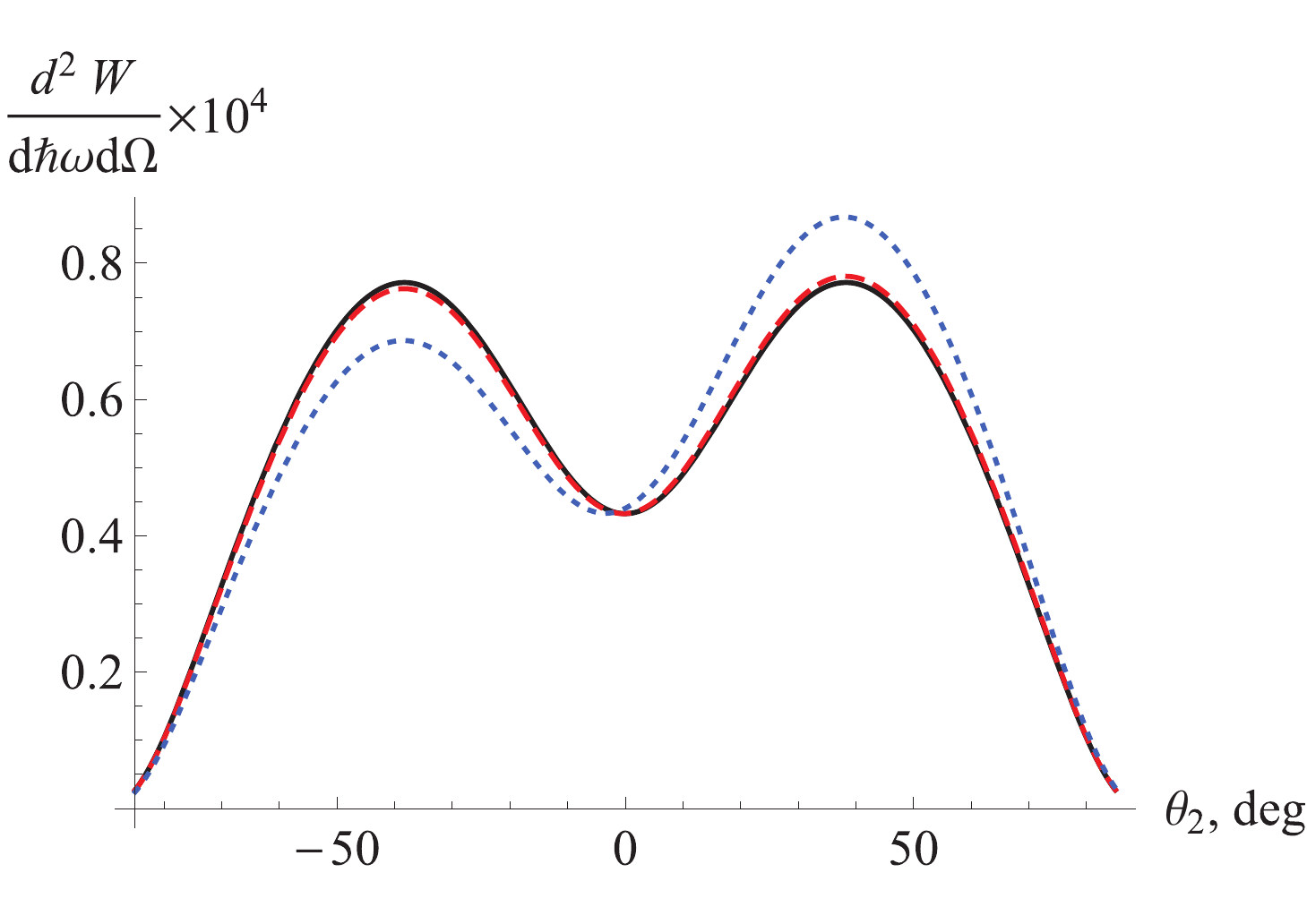}
\caption{(Color online.) Distribution in $\theta_2$ of the forward TR (left plot) and backward TR (right plot) for the benchmark 
case ($\alpha = 70^{\circ}, \bar\theta_1 = -40^{\circ}, \gamma = 1.59, \hbar\omega = 5$ eV) at $\ell = 0$ (solid black curve),  
$1000$ (dashed red curve) and $10000$ (blue dotted curve). The first model (see Sec.~\ref{method_1}) is used.}
   \label{fig-bench}
\end{figure}

Note that the $\theta_2$-distribution becomes strongly distorted at $\ell \sim 10^4$, which is consistent 
with the parameter $x_\ell \sim 10^{-5}\cdot \ell$ governing the magnitude of the left-right asymmetry.
For $\ell < 10^3$, the asymmetry is not easily discernible by eye, and it should be extracted via (\ref{asymmetry}).
Its value is shown in Fig.~\ref{fig-bench-l-dep}. As expected, it shows a nearly perfect proportionality to $\ell$.

\begin{figure}[!htb]
   \centering
\includegraphics[width=7cm]{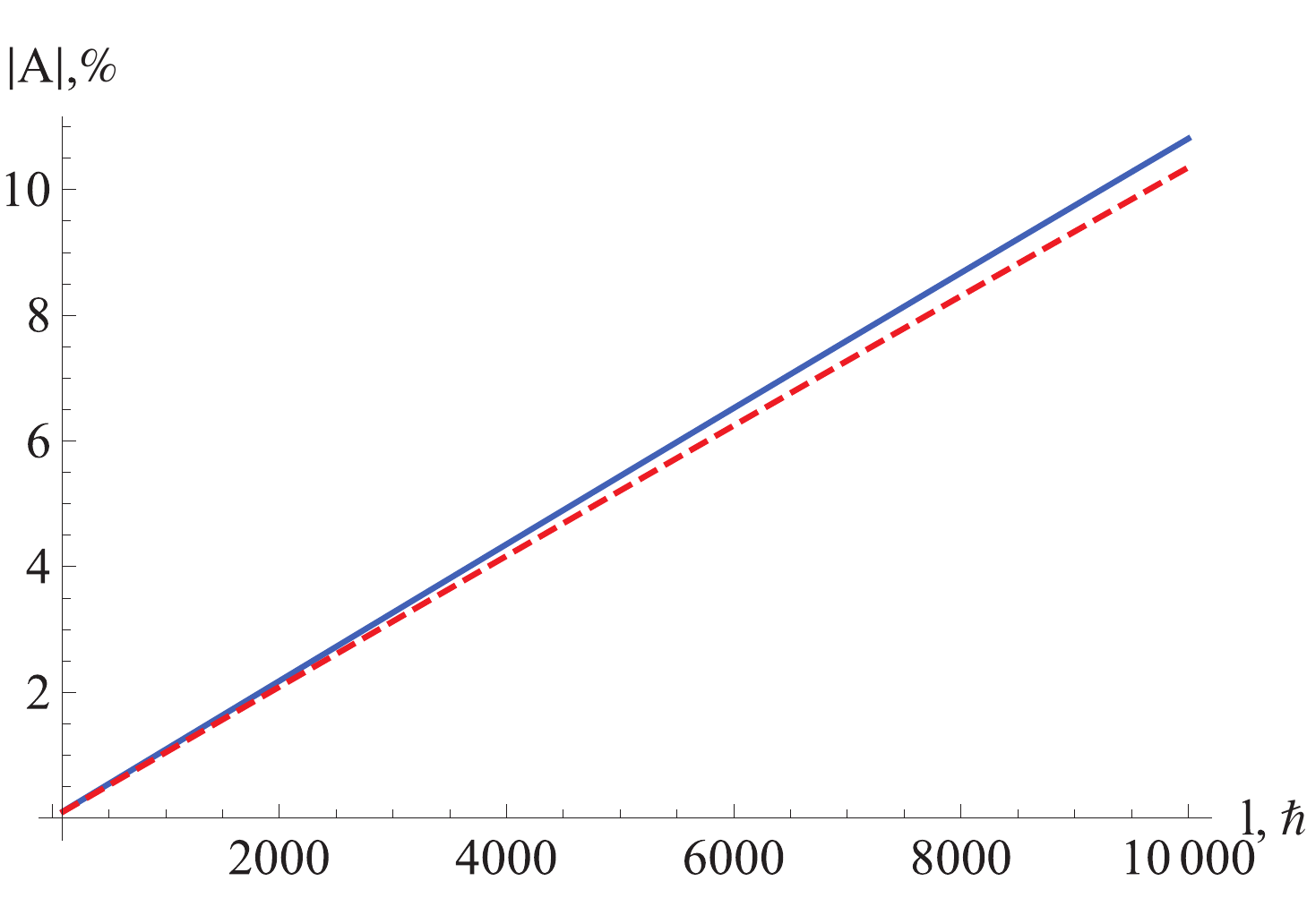}\hspace{1mm}
\includegraphics[width=7cm]{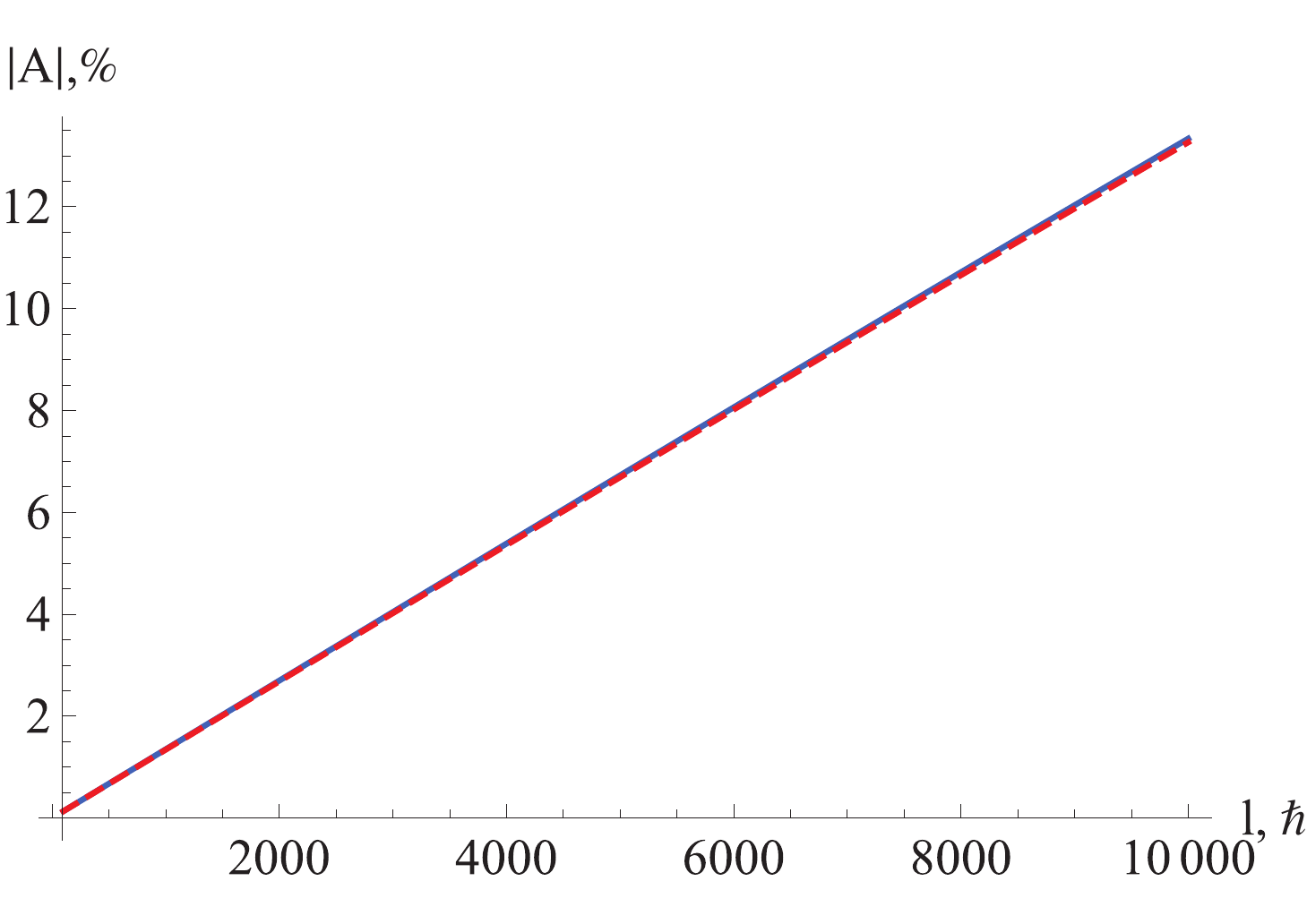}
\caption{(Color online.) The magnitude of the asymmetry $A$ in the benchmark case ($\bar\theta_1 = - 40^{\circ}$) as a function of $\ell$ for $\hbar\omega = 5$ eV (left) and $\hbar\omega = 10$ eV (right). The blue solid and the red dashed lines correspond to the forward and backward TR, respectively.}
   \label{fig-bench-l-dep}
\end{figure}

\begin{figure}[!htb]
   \centering
\includegraphics[width=7cm]{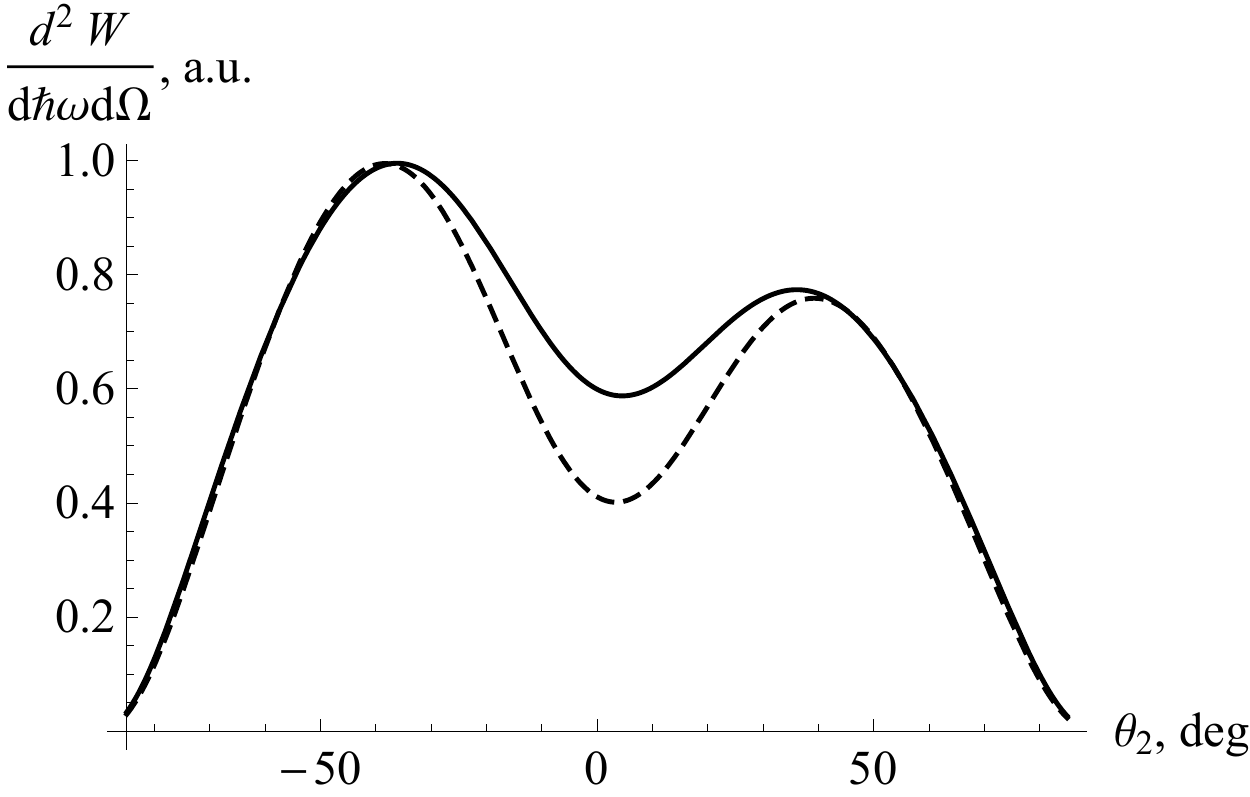}
\caption{Comparison of predictions for both models being used for the forward TR in the benchmark case and for $\ell = 10000$: the solid line corresponds to the first model (Sec.~\ref{method_1}), the dashed line corresponds to the second model (Sec.~\ref{method_2}). The values of the asymmetry agree within the accuracy better than $10$\%.}
   \label{Compare}
\end{figure}

A comparison of the two calculation methods being used is presented in Fig.~\ref{Compare} for the benchmark case. 
The difference between the predictions in the small-angle region does not affect the asymmetry values.

Finally, using the absolute value of the emitted energy distribution shown in Fig.~\ref{fig-bench3D} and in Fig.~\ref{fig-bench},
one can estimate that the average number of emitted UV photons (say, in the range of $3-10$ eV) per one incident electron
is $n_\gamma/n_e \sim {\cal O}(10^{-5}$--$10^{-4})$. For a current of $1$ nA it converts to ${\cal O}(10^{5}$--$10^{6})$ TR photons per second.

We would like to emphasize that the energies of vortex electrons achieved in electron microscopes so far do not surpass $300$ keV. 
Drawing an analogy with the ``pure'' charge TR, one could expect that the effect of interest would be detected much more easily in ultrarelativistic electrons. 
In fact, this is not the case as the electron energy dependence of the interference term $dW_{e\mu}$ is governed by the factor $\mu \omega/\gamma$, 
as is seen e.g. from the Eq.~(\ref{Hout}). The characteristic frequency of the forward TR depends linearly on the Lorentz-factor for ultrarelativistic electrons \cite{GTs, G}, 
$\omega \sim \gamma \omega_p$, so the ratio $\mu \omega/\gamma$ almost does not depend on the electron energy when $\gamma \gg 1$. 
Nevertheless, at the optical/UV frequencies, which are the most convenient in practice, the asymmetry is quickly damped with the electron energy rise, 
making the electrons with the energies of $200-300$ keV optimal for the detecting the effect. 
Note that the (charge) optical TR was successfully detected from the electrons with the energies of $80$ keV and even lower \cite{Nonrel}.

\subsection{Dependences}

Next, we show in Fig.~\ref{fig-angles} how the $\theta_2$-distribution changes upon a variation of the incidence angle $\alpha$
and the detection angle $\theta_1$. One sees that the two-bump structure becomes more pronounced
for a grazing incidence ($\alpha$ close to $90^\circ$) and for larger negative values of $\theta_1$.
This is convenient for detection of the (backward) TR, as the photons are to be detected at large angles $\sim 100^\circ$
with respect to the electron beam.

\begin{figure}[!htb]
   \centering
\includegraphics[width=5.5cm]{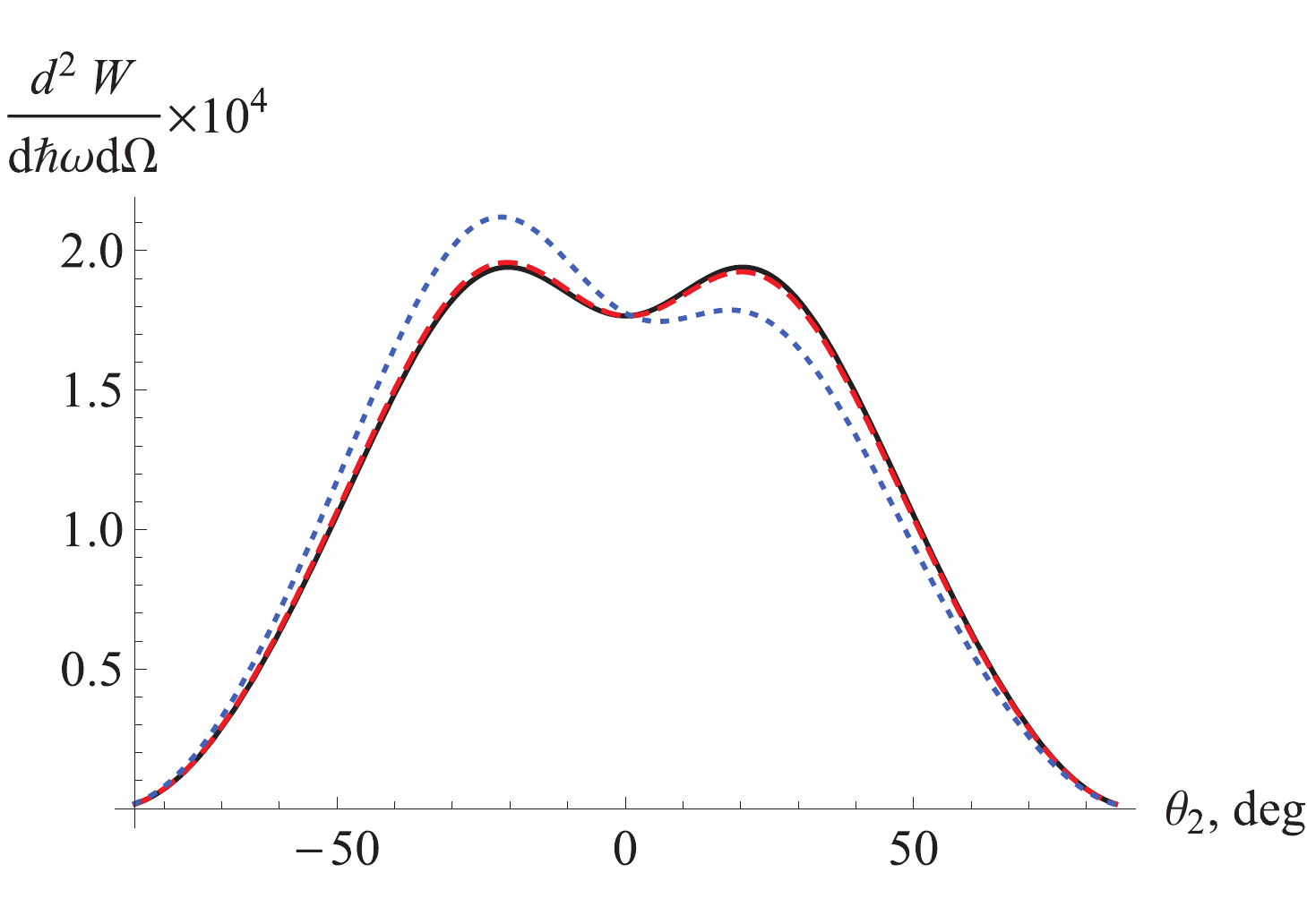}\hspace{1mm}
\includegraphics[width=5.5cm]{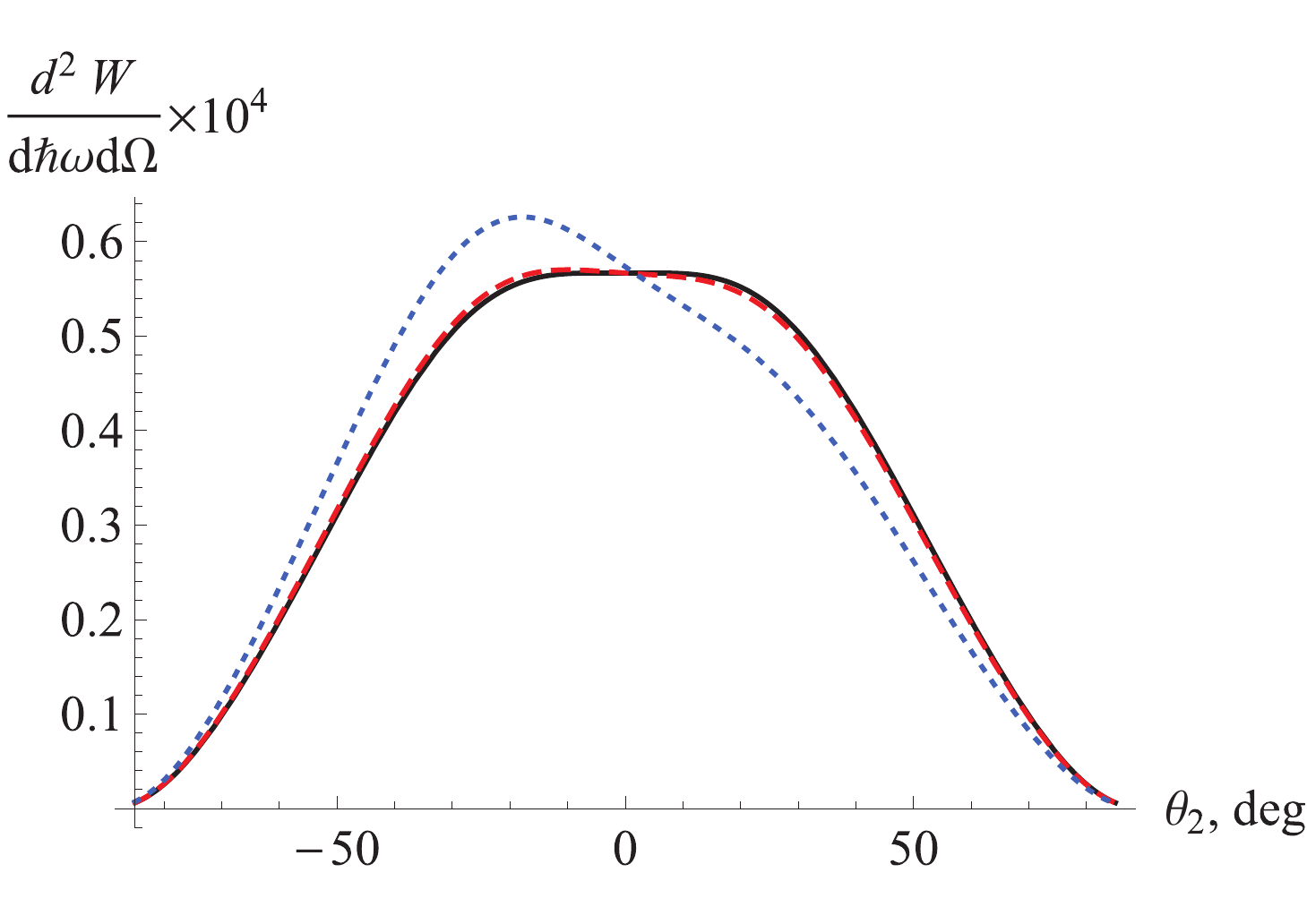}\hspace{1mm}
\includegraphics[width=5.5cm]{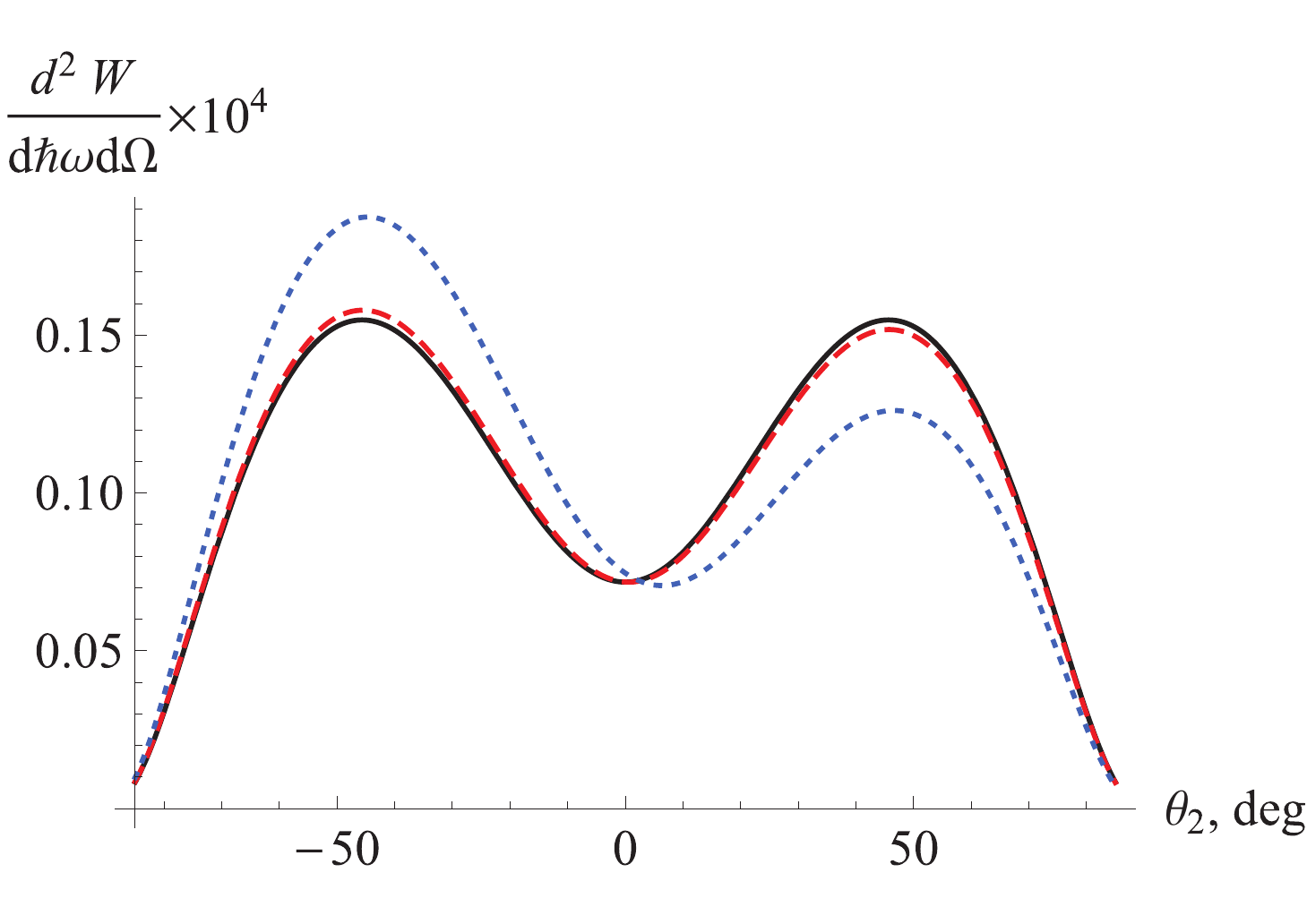}
\caption{(Color online.) Forward TR $\theta_2$-distributions for three other choices of the angles:
$\alpha = 70^\circ$, $\bar\theta_1=-20^\circ$ (left),
$\alpha = 80^\circ$, $\bar\theta_1=-40^\circ$ (middle),
$\alpha = 80^\circ$, $\bar\theta_1=-60^\circ$ (right).
In each case, $\ell = 0$ is shown by the solid black curve, $\ell=1000$ is shown by the dashed red curve, and $\ell = 10000$ is shown by the blue dotted one. 
The first model (see Sec.~\ref{method_1}) is used.}
   \label{fig-angles}
\end{figure}

\begin{figure}[!htb]
   \centering
\includegraphics[width=7cm]{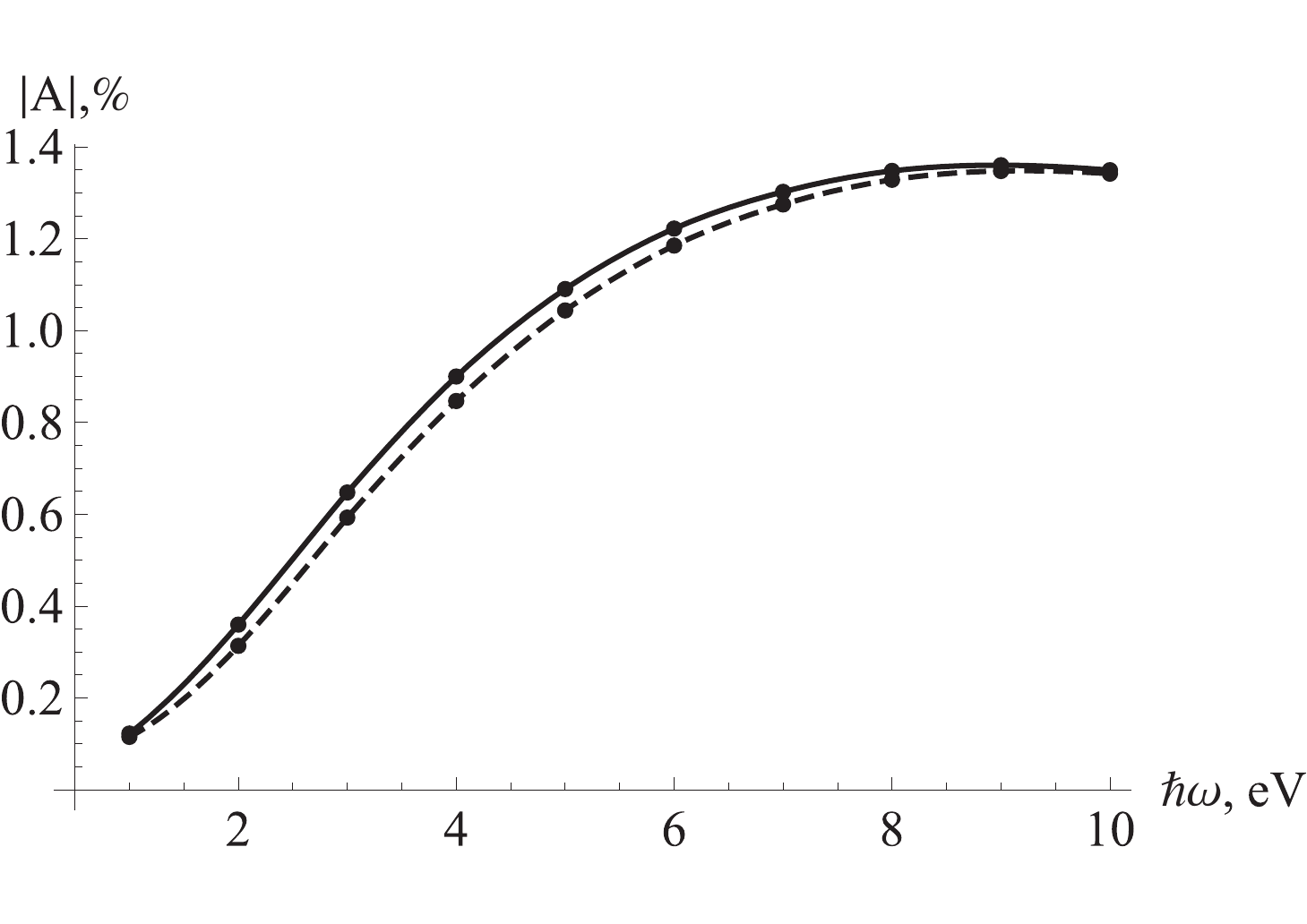}\hspace{5mm}
\includegraphics[width=7cm]{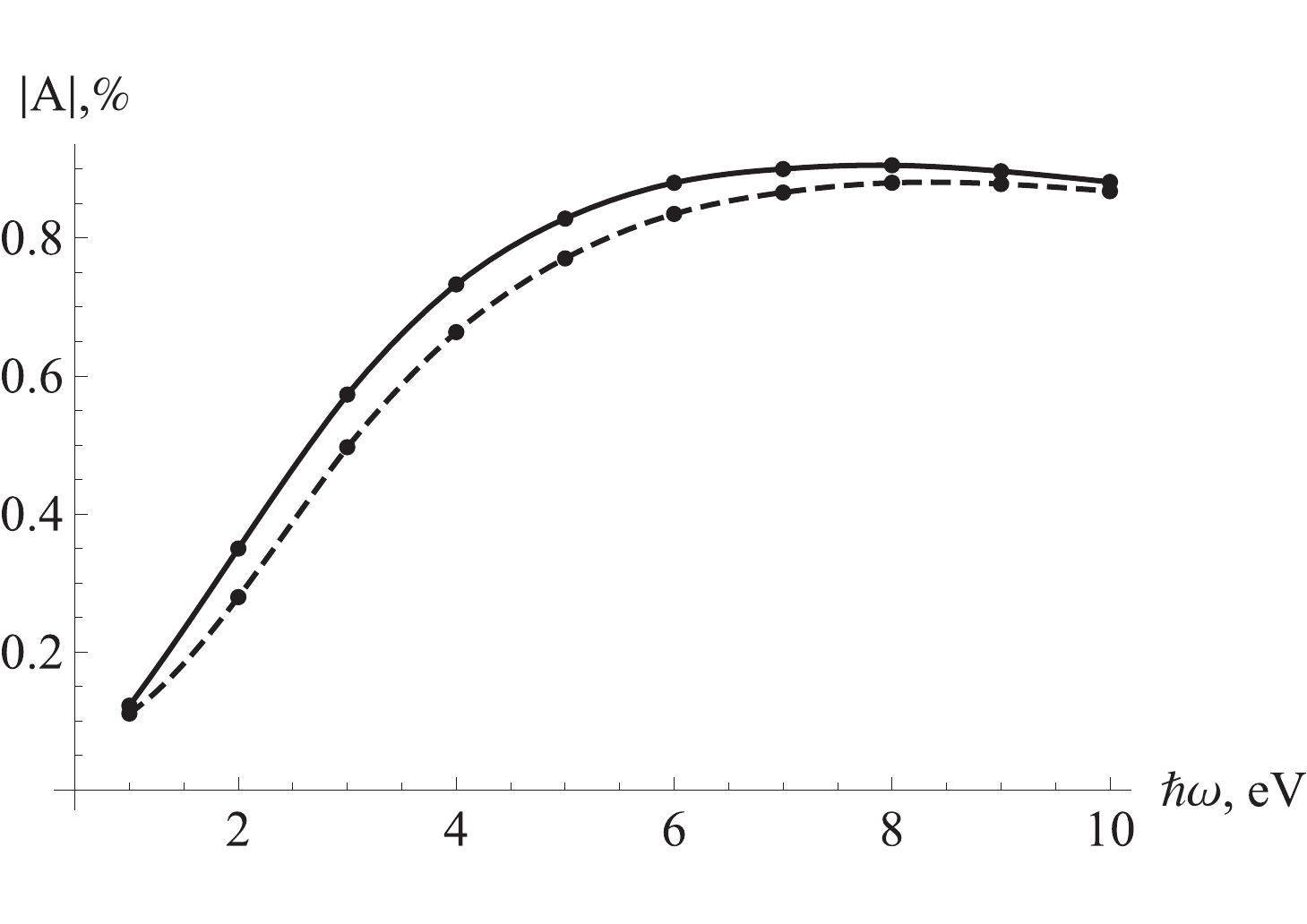}
\caption{The magnitude of the left-right asymmetry for the forward TR (solid line) and the backward TR (dashed line) 
as a function of the photon energy at the incidence angle $\alpha = 70^\circ$  and $\ell = 1000$. 
The two plots correspond to $\bar\theta_1=-40^\circ$ (left plot) and $\bar\theta_1=-20^\circ$ (right plot).}
   \label{fig-spectral}
\end{figure}
\begin{figure}[!htb]
   \centering
\includegraphics[width=7cm]{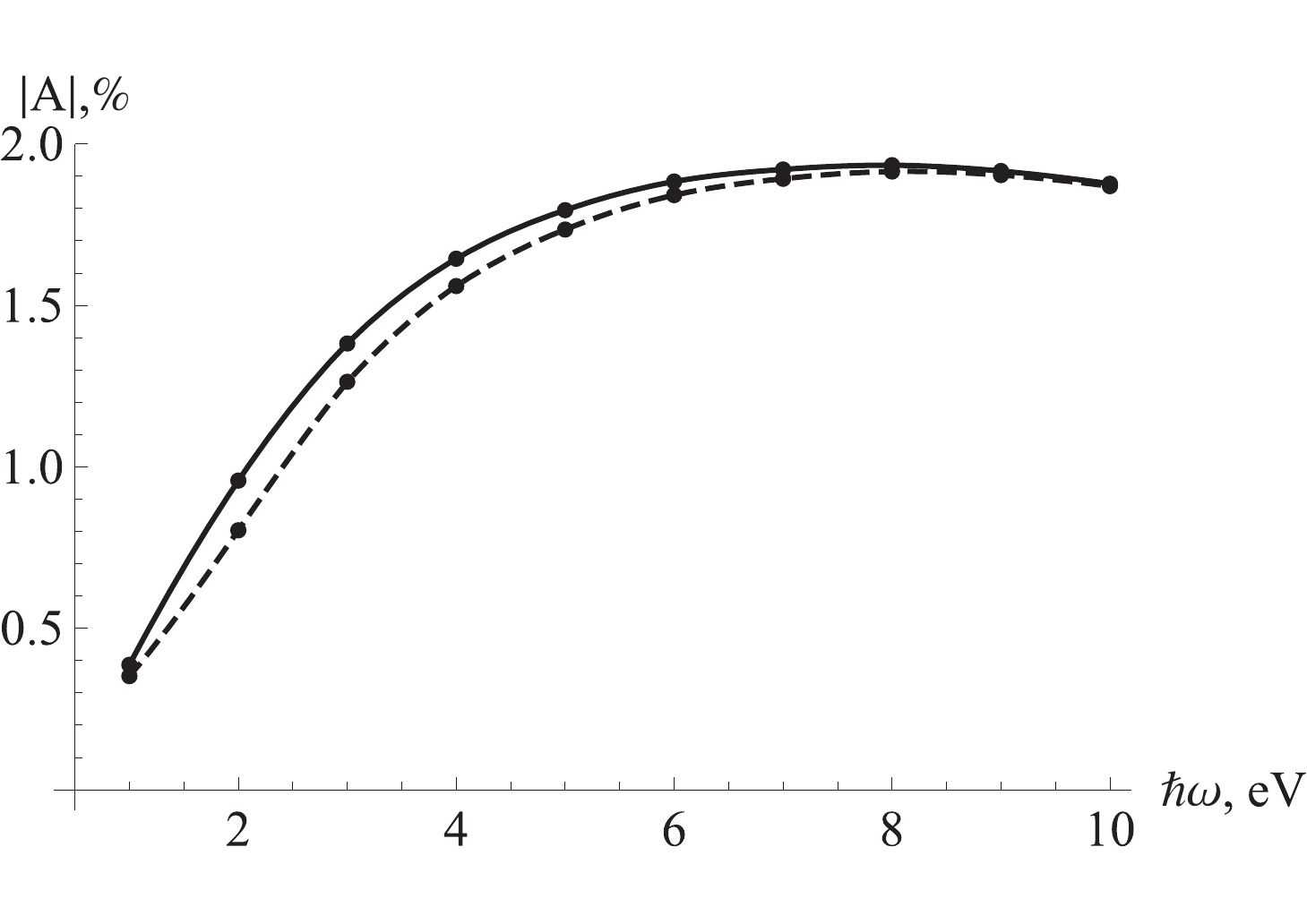}\hspace{5mm}
\includegraphics[width=7cm]{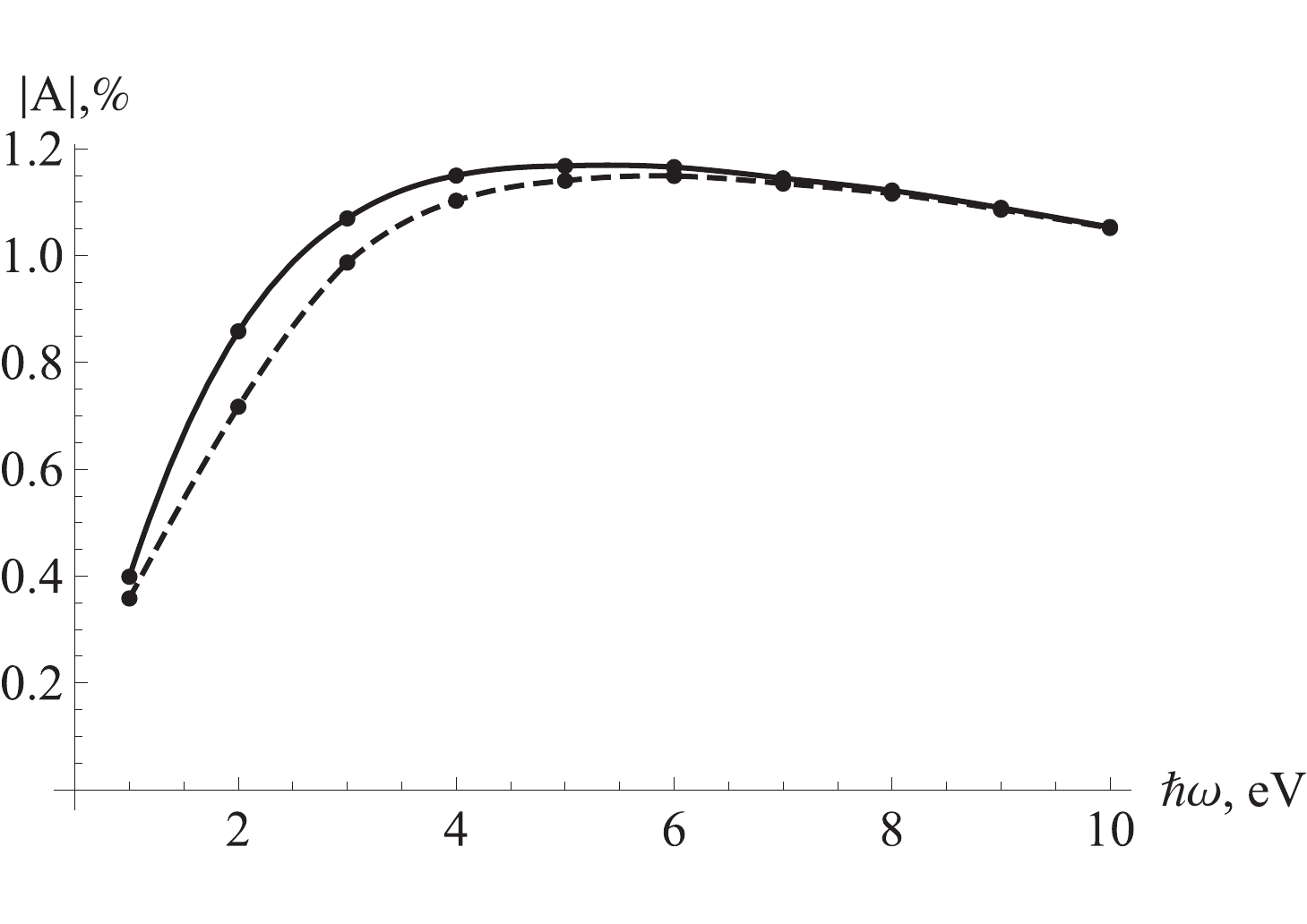}
\caption{The magnitude of the left-right asymmetry for the forward TR (solid line) and the backward TR (dashed line) as a function of the photon
energy at the incidence angle $\alpha = 80^\circ$  and $\ell = 1000$. 
The two plots correspond to $\bar\theta_1=-60^\circ$ (left plot) and $\bar\theta_1=-40^\circ$ (right plot).}
   \label{fig-spectral_2}
\end{figure}

\begin{figure}[!htb]
   \centering
\includegraphics[width=7cm]{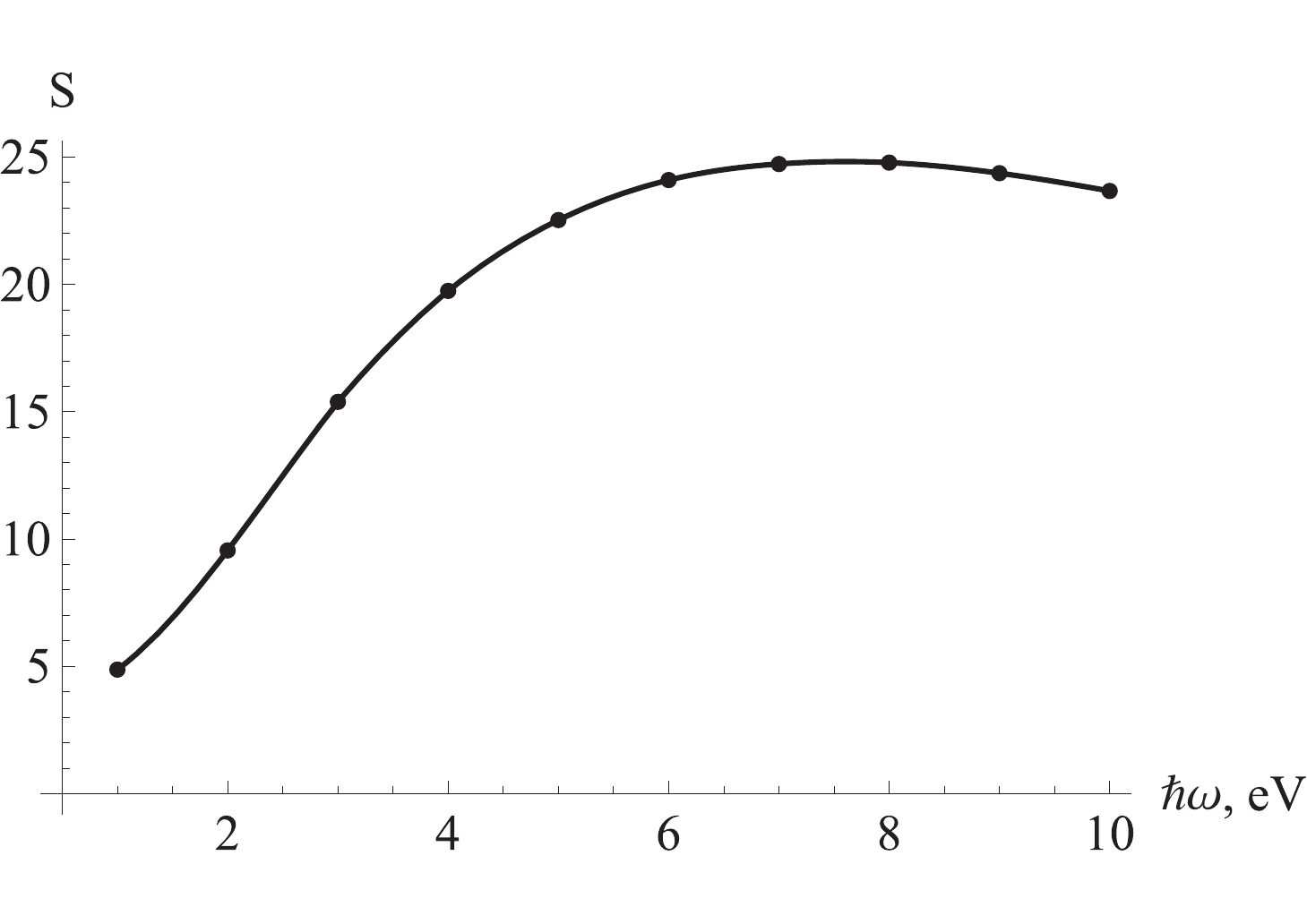}\hspace{5mm}
\includegraphics[width=7cm]{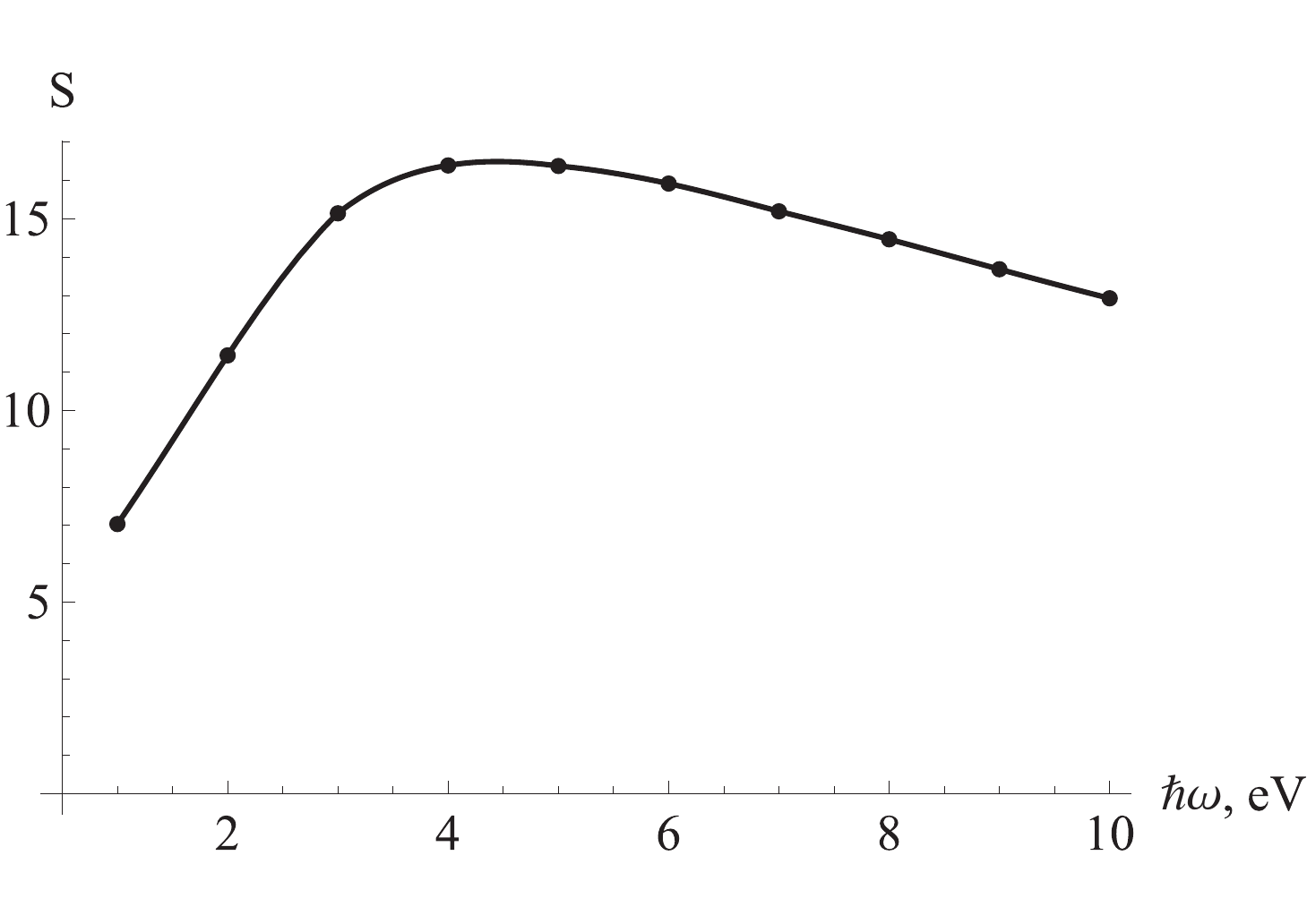}
\caption{Statistical significance of the backward TR left-right asymmetry for $\ell = 1000$, 
based on a statistics of $5\cdot 10^{12}$ electrons
and assuming the $10$\% quantum efficiency for the photon detector
(left panel: $\alpha = 70^{\circ}, \bar\theta_1 = -40^{\circ}$, right panel: $\alpha = 80^{\circ}, \bar\theta_1 = -60^{\circ}$).
Integration in $\theta_2$ goes over $\pm(10^\circ, 90^\circ)$.}
   \label{fig-statistical}
\end{figure}

The spectral dependence of the asymmetry is shown in Figs.~\ref{fig-spectral},~\ref{fig-spectral_2}.
Note that the initial rise $|A| \propto \hbar \omega$ is quickly tamed in the near UV region due to the permittivity frequency dispersion,
which makes this region best for detecting the effect. Note that we employ the simplest definition of the asymmetry 
(\ref{asymmetry}), whereas an alternative definition, (\ref{asymmetry_0}) with $f(\theta_2) = \sin(\theta_2)$, 
yields even larger values of $A$ (up to $\sim 1.2$ times the benchmark case values). 

To quantify the visibility of the asymmetry, we introduce its statistical significance $S$. It shows how the ``true'' extracted asymmetry
compares to a typical ``fake'' asymmetry, which might arise in a perfectly symmetric distribution due to a statistical fluctuation
in the photon counting statistics.
If the asymmetry is calculated according to (\ref{asymmetry_0}) with the weight function $f(\theta_2)$,
and if the total number of the incident electrons integrated over a certain time is $N_e$,
then we define the weighted total photon count $N_\gamma$ as follows:
\be
N_\gamma(\bar\omega) = N_e \int {d\omega \over \omega }\int d\Omega\, |f(\theta_2)| {d^2W \over d \omega d\Omega}\,.
\ee
The spectral integral here extends over a certain region centered as $\bar\omega$ (for the estimates below, we use $1$ eV-wide bins).
The left-right asymmetry of the counts is then
\be
\Delta N_\gamma(\bar\omega) = N_e \int {d\omega \over \omega }\int d\Omega\, f(\theta_2) {d^2W \over d \omega d\Omega}\,.
\ee
The expected mean value of the statistical fluctuation of $\Delta N_\gamma$ is $\sqrt{N_\gamma}$.
Therefore, the statistical significance is defined by
\be
S(\bar\omega)  = {\Delta N_\gamma(\bar\omega)  \over \sqrt{N_\gamma(\bar\omega)}}\,.
\ee
The true statistical significance of the count difference detected in an experiment will certainly be smaller due to 
systematic uncertainties. However $S(\bar\omega)$ still gives a good idea of the needed integration time and
of the $\bar\omega$ region optimal for the asymmetry detection.
We plot this quantity in Fig.~\ref{fig-statistical} for the statistics of $5\cdot 10^{12}$ incident electrons,
which corresponds to the integration time of $\approx 15$ min at the current of $1$ nA.
The quantum efficiency of the photon detector is assumed to be $10$\%.
We see that at these parameters the asymmetry should be very visible, and the optimal
frequency range is near UV. The values shown in Fig.~\ref{fig-statistical} correspond to $f(\theta_2) = \sgn (\theta_2)$. 
The analogous choice, $f(\theta_2) = \sin (\theta_2)$,  
yields slightly lower values of $S$ ($\sim 0.8$ of the values shown in Fig.~\ref{fig-statistical}). 

As for the sensitivity of the results to the values of permittivity, we only mention here that 
this dependence is rather weak provided the substance under consideration has prominent absorption, $\varepsilon^{\prime\prime}$. 
We obtain asymmetries of the same order of magnitude by varying $\varepsilon'$ and $\varepsilon''$, 
see Table \ref{table}. Therefore, we expect a similar visibility for other metals.
\begin{table}
\caption{The asymmetry values for the backward TR in the benchmark case with $\ell = 1000, \omega = 5$ eV as a function of permittivity with aluminium taken as a benchmark: $\varepsilon_{\text{Al}} (\omega) \approx - 8.38 + i\, 1.05$ \cite{Rakic}.} \label{table}
\begin{center}
\begin{tabular}{|c|c|c|c|}
\hline
A,\% &$\varepsilon^{\prime\prime}_{\text{Al}}$& $0.2 \varepsilon^{\prime\prime}_{\text{Al}}$& $5 \varepsilon^{\prime\prime}_{\text{Al}}$ \\
\hline
$\varepsilon^{\prime}_{\text{Al}}$ &$1.0$ (Al) & $1.1$ & $0.9$ \\
\hline
$0.2 \varepsilon^{\prime}_{\text{Al}}$ & $0.7$ & $0.7$ & $0.7$ \\
\hline
$5 \varepsilon^{\prime}_{\text{Al}}$ & $0.9$ & $0.9$ & $0.9$ \\
\hline
\end{tabular}
\end{center}
\end{table}

\section{Discussion}\label{section-discussion}

\subsection{Experimental feasibility}

In this section we provide some rough estimates which show that the effect
can in principle be observed with the existing technology and requires only moderate
adjustments to the electronic microscopes currently used for the vortex electron generation.

The key issue enabling the observations we suggest is creation of the vortex electrons with a large OAM.
The figure of merit here is not the largest OAM by itself, but the OAM value at the first diffraction peak (the higher order peaks
are strongly suppressed in intensity). 
The maximal value achieved so far is $25$ \cite{vortex-experimental-mcmorran}; a tenfold increase of this value is highly desirable.
This will certainly pose a challenge in manufacturing the appropriate diffraction gratings,
but these values seem to be within technological limits. Indeed, a typical aperture available at the position of the condenser lens is of 
the order of hundred microns,
while the smallest features which can be accurately etched in a grating are of tens of nanometers.

It might also be possible to create the very high OAM vortex electrons using the recently demonstrated
technique of electron scattering on an effective magnetic monopole, \cite{verbeeck-monopole}.
In this experiment, the ring-shaped nonvortex electron wave passes through an open end of
a magnetic whisker or a nanoscale solenoid, whose field is well approximated locally by a magnetic monopole field, 
and it acquires vorticity.
The OAM value is determined by the effective magnetic charge of the monopole, which can in principle
be made very large.

It must be stressed that our suggestion does not require the vortex electrons to be in a state
of definite value of $\ell$. Quite to the contrary, the OAM can be spread over a certain
rather broad range, and the effect will still be there.
Even if the transverse profile of the electron state becomes distorted, this does not have any sizable
effect on the asymmetry because all transverse shifts remain much smaller than $\lambda$.
This makes our predictions robust against imperfections of the experimental method of generating the high-OAM vortex electrons.

A similar conclusion holds for another distortion effect. It can be expected that higher order
phase vortices are inherently unstable. Upon propagation in a magnetic lense system with stray fields, 
they might split into a compact ``cloud'' of vortices of topological order one.
This possibility however does not affect the predicted asymmetry if the ``cloud'' stays compact, $\ll \lambda$.

The second delicate issue is the alignment of the photon detectors.
We propose to place two identical large aperture detectors symmetrically on the two sides
of the electron incidence plane. They do not even have to be pixelated, because the quantity
to be measured is the asymmetry between the left and right detectors.
All instrumentation alignment should be performed with a relative accuracy better than the estimated asymmetry.

If achieving an accurate symmetric alignment proves difficult, one can then fix the instrumentation and 
simply change the sign of the OAM of vortex electrons. This can be done by tilting the grating without
any mechanical manipulation to the target or to detectors. One should then observe the sign change in the asymmetry.

The third issue concerns the expected energy of TR. 
Using our estimates of $n_\gamma \sim {\cal O}(10^{-4})$ photons per electron and 
taking a current of 1 nA, which is easily achievable
in microscopes producing vortex electrons, one can expect about $10^5$ photons per second
detected by photocathodes with quantum efficiency 10\%.
A sufficiently long integration time will lead to $10^8$ photons, and with this statistics 
a left-right asymmetry of the order $A \sim 0.1\%$ can be reliably detected.

Finally, let us comment on coherence issues.
The coherence condition for radiation (focusing the electron beam to spots much smaller than the wavelength of the emitted
light) can be easily achieved with existing devices. 
Focusing a vortex beam with small $\ell$ to angstrom scale spots has been demonstrated \cite{verbeeck-2},
and one can expect that focusing electrons with $\ell = 1000$ to submicron scales should also be feasible.

Longitudinal extent of the individual electron wave function should also be below optical wavelength,
which means that the longitudinal self-correlation length of the electron beam should not be too good.
This can be cast in the form of the requirement that the monochromaticity of the electrons
should be {\em worse} than a few eV.

\subsection{The effect in other forms of radiation}

In this paper we have discussed possibilities for detecting the interference term $dW_{e\mu}$ in Cherenkov radiation 
and transition radiation. These phenomena represent, in fact, two particular cases of the general process of polarization radiation. 
Rather simple considerations allow one to estimate the magnitude of similar effects in other processes like, for instance, 
diffraction radiation and Smith-Purcell radiation. Indeed, for an observer located far enough from the target (in the wave zone), 
the radiation arises as a result of a distant collision of a (vortex) electron with a pointlike dipole moment, $d (\omega)$. 
Irrespectively of the target shape, the radiation field in the wave zone is (integration is over the target volume)
\be
{\bm H}^{PR} \propto {\bm e}_k \times \int \limits_V  d^3r\, {\bm d} ({\bm r}, \omega) e^{-i ({\bm k}{\bm r})} \propto {\bm e}_k \times {\bm d}(\omega)
\ee
since in the dipole approximation ${\bm j}^{pol} = -i\omega {\bm d}$. Roughly speaking, it is the explicit expression for the dipole moment ${\bm d}$ 
that only makes difference in different types of polarization radiation. 
As a result, the product 
$$
{\bm e}_k \cdot [{\bm \mu} {\bm d}]
$$
or ${\bm e}_k \cdot [{\bm e}_u {\bm d}]$ (since ${\bm \mu} \parallel {\bm u}$) will govern the effect (with ${\bm d}$ instead of the normal ${\bm n}$). 
From this, it is immediately clear that the interference effect is absent for Cherenkov radiation, even for arbitrary complex $\varepsilon (\omega)$, 
for transition radiation at the normal incidence, and also for diffraction radiation when the particle moves \textit{nearby} a metallic foil, 
perpendicular to the surface, but does not intersect it (for a detailed description see e.g., \cite{DR}). 
The effective dipole moment in most cases of practical interest is perpendicular to the target surface \cite{JETP}, 
so in all the geometries mentioned ${\bm d} \parallel {\bm u}$. On the contrary, the effect will exist when a particle moves \textit{obliquely} 
with respect to the target surface in the diffraction radiation problem or even when it moves nearby a metallic grating as in Smith-Purcell radiation. 
In all these geometries one could expect the same angular asymmetry, which should increase as the angle between ${\bm d}$ and ${\bm u}$ grows, 
and its numerical value will be of the same order as in the TR case. Finally, one can note that when dealing with other types of PR 
the actual dielectric properties of the target materials and the frequency dispersion are highly important, 
and they can be taken into account with the approach developed in Ref.~\cite{JETP}.

One could also mention that if vortex electrons with the high values of OAM were created, detection of the radiation asymmetry could serve 
as a diagnostic tool allowing one to obtain the value of $\ell$ of the beam. Such a diagnostics could be done noninvasively 
by using diffraction radiation from a rectangular plate instead of transition radiation. The former has the same angular distributions 
in the $\theta_2$-plane as in the TR case considered here, but the beam characteristics stay undisturbed during the emission process (see e.g. \cite{DR}).

Along with the OAM-induced effects in PR discussed in the present paper, there is also the possibility to study similar effects 
in radiation processes in external high-intensity electromagnetic fields. Indeed, the magnitude of the spin effects in such a field is governed by 
the Lorentz-invariant ratio $E^{\prime}/E_{cr}$ (see e.g., \cite{Br}; here, $E^{\prime}$ is the electric field strength in the particle rest frame 
and $E_{cr} = 1.3 \cdot 10^{18}\, \text{V}\,\text{m}^{-1}$ is the ``critical'' Sauter-Schwinger value). 
Its counterpart for the OAM-induced magnetic moment effects is 
\be
{({\bm \mu} {\bm H}^{\prime}) \over m c^2} = {\ell H_z^{\prime} \over 2H_{cr}} \gg {H_z^{\prime} \over H_{cr}}\,,\quad
​H_{cr} = 4.4\,\cdot 10^{13}\, G\,,
\ee
which is also a Lorentz-invariant expression (here $H_z^{\prime}$ is the magnetic field projection onto the propagation direction in the electron rest frame). 
This means, roughly speaking, that the requirements for the field strength to make the magnetic moment effects in radiation observable become much more relaxed 
for vortex electrons with $\ell \gg 1$.
​Accordingly, if such electrons with $\ell \sim 100-1000$ were accelerated up to the energies of $100 \text{MeV} - 1 \text{GeV}$, 
this would allow one to study effects analogous to the spin effects in radiation of the non-vortex electrons (see e.g, \cite{Br}). 
Such an acceleration seems to be feasible, at least in principle, with the novel technique recently demonstrated in Ref.~\cite{verbeeck-monopole} 
(see also discussion in Ref.~\cite{K-2012}).

\section{Conclusions}

Recently created vortex electrons carrying large orbital angular momentum $\ell$ and, therefore,
a large OAM-induced magnetic moment are an ideal tool to investigate 
influence of the magnetic moment on various forms of polarization radiation.
This influence has been discussed theoretically since long ago but up to now has never been studied experimentally. 
As the magnetic moment contribution is parametrically suppressed by the small parameter $x_\ell = \ell \hbar\omega/E_e$,
one can hope to detect it only via its interference with the charge contribution.
This interference can be extracted via an angular asymmetry,
but even here one must strive for largest achievable $\ell$.

In this paper, we investigated this effect for different types of polarization radiation.
We showed the absence of the interference term for Cherenkov radiation,
studied in detail the interference and the asymmetry for transition radiation,
and commented on possibility to observe this effect for other forms of PR.
In particular, we argued that for $\ell = 100-1000$, the asymmetry in TR can be of the order of $0.1\%-1\%$, 
which could be measurable with existing technology. 
Simultaneously, it offers a novel method of measuring large OAM
in electron vortex beams. 

\begin{acknowledgments}
I.P.I. acknowledges grants RFBR 11-02-00242-a and RF President grant for scientific schools NSc-3802.2012.2.
D.V.K. acknowledges grants of the Russian Ministry for Education and Science within the program ``Nauka'' and Nos. 14.B37.21.0911, 14.B37.21.1298. 
The authors are grateful to V.~G.~Bagrov and A.~A.~Tishchenko for useful comments and also to J.~Verbeeck and members of his team at Antwerpen University 
for discussions on experimental feasibility of the proposed measurement.
\end{acknowledgments}

\end{document}